\documentclass[utf8]{frontiersSCNS} 

\usepackage{url,hyperref,lineno,microtype,subcaption}
\usepackage[onehalfspacing]{setspace}
\usepackage{graphicx}
\usepackage{xcolor}
\usepackage{natbib}
\usepackage{txfonts}
\usepackage{color} 
\usepackage{multirow}
\usepackage{rotating}
\usepackage{lscape}
\usepackage{enumerate}
\usepackage{hyperref}
\usepackage[T1]{fontenc}
\usepackage{lmodern}
\hypersetup{colorlinks=true, linkcolor=red, citecolor = blue, filecolor=cyan,  urlcolor=magenta,}

\newcommand{\aap}{    {\it Astron. Astrophys.}}
\newcommand{\aaps}{   {\it Astron. Astrophys. Suppl.}}

\newcommand{\apj}{    {\it Astrophys. J.}}
\newcommand{\apjl}{   {\it Astrophys. J. Lett.}}

\newcommand{\solphys}{{\it Solar Phys.}}

\chardef\us=`\_

\def\keyFont{\fontsize{8}{11}\helveticabold }
\def\firstAuthorLast{Patel {et~al.}} 
\def\Authors{Ritesh Patel\,$^{1,2,*}$, Megha A.\,$^{1}$, Arpit Kumar Shrivastav\,$^{1,2,3}$, Vaibhav Pant\,$^{2,*}$, M. Vishnu\,$^{1}$, Sankarasubramanian K.\,$^{1,4,5}$ and Dipankar Banerjee\,$^{1,2,4}$}
\def\Address{$^{1}$Indian Institute of Astrophysics, 2nd Block Koramangala, Bangalore 560034, India \\
$^{2}$Aryabhatta Research Institute of Observational Sciences, Nainital 263001, India \\
$^{3}$Indian Institute of Science, CV Raman Road, Bangalore 560012, India \\
$^{4}$Center of Excellence in Space Science, IISER Kolkata, Kolkata 741246, India \\
$^{5}$UR Rao Satellite Centre, Old Airport Road, Bangalore 560017, India 
}
\def\corrAuthor{Ritesh Patel, Vaibhav Pant}

\def\corrEmail{ritesh.patel@iiap.res.in, vaibhav.pant@aries.res.in}

\begin{document}
\onecolumn
\firstpage{1}

\title[Characterizing VELC Spectral Channels]{Characterizing Spectral Channels of Visible Emission Line Coronagraph of Aditya-L1} 

\author[\firstAuthorLast ]{\Authors} 
\address{} 
\correspondance{} 

\extraAuth{}

\maketitle

\begin{abstract}

\section{}
Aditya-L1 is India’s first solar mission with Visible Emission Line Coronagraph (VELC) consisting of three spectral channels taking high-resolution spectroscopic observations of the inner corona up to 1.5~R$_\odot$ at 5303 \AA, 7892 \AA, and 10747 \AA. In this work, we present the strategy for the slit-width optimization for the VELC using synthetic line profiles by taking into account the instrument characteristics and coronal conditions for log(T) varying from 6 to 6.5. The synthetic profiles are convolved with simulated instrumental scattered light and noise to estimate the signal-to-noise ratio (SNR), which will be crucial to design the future observation plans. We find that the optimum slit width for VELC turns out to be 50 $\mu$m providing sufficient SNR for observations in different solar conditions. We also analyzed the effect of plasma temperature on the SNR at different heights in the VELC field-of-view for the optimized slit-width. We also studied the expected effect of the presence of a CME on the spectral channel observations. This analysis will help to plan the science observations of VELC in different solar conditions.

\tiny
 \keyFont{ \section{Keywords:} Corona, Coronagraph, Spectroscopy, Emission lines, Instrumentation} 
\end{abstract}

\section{Introduction}

The observations of the inner corona ({up to} 3 R$_\odot$) in white-light and emission lines have been made during the total solar eclipses over the years yielding a detailed description of the corona \citep{Baumbach37, Hulst50, Habbal2007ApJ, Habbal2010ApJ, Habbal_2014, Boe2018Freezin}. During the total solar eclipses, spectroscopic investigations utilizing the emission lines  detected the presence of oscillations and fast magnetohydrodynamic (MHD) waves in the solar atmosphere \citep{Singh1997SoPh, Pasachoff2002SoPh, Sakurai2002SoPh, Singh2011SoPh, Samanta2016SoPh}.
As eclipses last for a couple of minutes, regular observations of the inner corona will help in improving our understanding of the solar atmosphere. It may also shed light on the development of small and large scale transients that lead to severe space-weather. The availability of observations in emission lines such as H$\alpha$ 6563 \AA, Fe IX 4359 \AA, Fe X 6374 \AA, Fe XI 7892 \AA, Fe XIII 10747 \AA, Fe XIV 5303 \AA, and Ni XV 6702 \AA\  can provide useful thermodynamic diagnostics of different processes occurring in the solar corona \citep{Habbal2011ApJ}. It should be noted that current ground- and space- based instruments lack coronal observations in majority of these wavelengths. The Coronal Multi-channel Polarimeter (CoMP) instrument \citep{COMP2008SoPh} in Mauna Loa Solar Observatory (MLSO) provided regular spectro-polarimetric observations of the inner corona up-to 1.5 R$_\odot$ at 10747 \AA, and 10798 \AA\ until 2018.  The detection of Alfv\'enic waves in the solar corona using the CoMP data provided support for wave based models for coronal heating \citep{Tomczyk2007Sci, Morton2015NatCo, Morton2016ApJ, Morton2019NatAs}.
The observations by this instrument also provided the first global magnetic map of the solar corona \citep{Yang2020Sci}.
Another ground based instrument, Norikura coronagraph, located at Norikura, Japan provided spectroscopic observations of the off-limb corona in emission lines corresponding to Fe X, Fe XI,, Fe XIII and Fe XIV. The emission line-intensity ratios give an estimate of the temperature while the line width could be used to calculate the thermal and non-thermal structure of the emitting plasma. Such high resolution spectroscopic observations from Norikura and CoMP revealed the thermal and non-thermal variations in coronal structures and velocities, complex variations in line-intensity ratios inferring the distribution of multi-thermal plasma and turbulence in these structures \citep{Singh_2003, Singh2004ApJ, Singh2006SoPh, Singh_2006, Krishna2013ApJ, McIntosh2012ApJ, Morton2016ApJ, Fan2018SoPh, Tiwari2019ApJ, Pant2019ApJ}.

The variations in the temperature and non-thermal velocity in coronal structures may give an insight of the processes involved in heating the corona and acceleration of solar wind. Therefore, continuous spectroscopic monitoring of the solar corona in such emission lines is required. Among the existing space-based sprectrograph instruments, EUV Imaging Spectrometer (EIS) on Hinode takes the observations of the solar corona and upper transition region in the wavelength range of 170 – 210 \AA\ and 250 – 290 \AA\ \citep{EIS2007SoPh}. Another spacecraft, Interface Region Imaging Spectrograph (IRIS) takes simultaneous imaging and spectroscopic observations of the photosphere, chromosphere, transition region and corona in three pass-bands of 1332--1358 \AA, 1389--1407 \AA, and 2783--2834 \AA\ \citep{IRIS2014SoPh}. The Spectral Imaging of the Coronal Environment (SPICE) instrument \citep{SPICE2020} on-board the recently launched Solar Orbiter is an imaging spectrometer capable of observing the corona in extreme ultraviolet (EUV) pass-bands of 70.4 nm -- 79.0 nm and 97.3 nm -- 104.9 nm. It should be noted that EIS, IRIS, and SPICE perform the observation confined to a small field of view (FOV).

Visible Emission Line Coronagraph (VELC) on-board Aditya-L1 \citep{ADITYA2017} will take simultaneous imaging and spectroscopic observations of the inner solar corona in three visible and one infra-red (IR) passbands from 1.05-1.5 R$_\odot$ \citep{Singh13, VELC17, IAUS2017}. VELC is equipped with a multi-slit spectrograph to study the solar corona with high spatial and temporal resolutions in the three emission lines centered at 5303 \AA\ (Fe XIV), 7892 \AA\ (Fe XI), and 10747 \AA\ (Fe XIII) mostly used during the eclipse observations. Continuous observations provided by VELC will be helpful to attain its scientific objectives including the diagnostics of the coronal plasma for temperature, and velocity thereby understanding the process of coronal heating and acceleration of the solar wind. Moreover, the spectro-polarimeteric capability of VELC using 10747 \AA\ emission line will be helpful to directly estimate the magnetic fields of the solar corona. Therefore, it is necessary to understand the performance of the instrument beforehand to plan the observations after launch.

In this article we present the characterization of the VELC spectral channels using synthetic spectra generated using {CHIANTI 8.0} for different coronal conditions taking in to account for instrument parameters. In Section \ref{sec:synthSpectra} we present the process involved in synthesizing the spectra and converting it to simulated observations. We present a detailed analysis and the results including the optimization of slit-width of VELC followed by the instrument performance for different conditions in Section \ref{sec:results}. Section \ref{sec:summary} summarizes the analysis followed by a discussion.

\section{Synthesizing Spectra}
    \label{sec:synthSpectra}

As discussed before VELC will take spectroscopic observations of the inner corona at three emission wavelengths centered at 5303 \AA, 7892 \AA, and 10747 \AA. We used {CHIANTI 8.0} atomic database \citep{CHIANTI1997, Zanna2015A&A} to generate emission spectra for these three ionisation states of iron. {It should be noted that the mechanism responsible for emission corona is atomic transitions unlike scattering of photospheric light in case of K and F corona. Also, it has been found that the F corona dominates beyond 2.5-3 R$_\odot$ \citep{Morgan07} which is beyond VELC field of view (FOV) of the spectroscopic channels. The emission lines under consideration are about 100 times brighter than the background K continuum \citep{Stix02}. We have included the continuum while preparing the synthetic spectra with CHIANTI.} The contribution of the instrument to the full width half maximum (FWHM) of the spectra is calculated using the following relation:

\begin{equation}
\label{eq:fwhm}
    \mathrm{FWHM_{instr}} = \sqrt{\Bigg (\frac{\mathrm{dispersion}}{\mathrm{pixel \; scale}} \times \mathrm{slit \; scale} \Bigg )^{2} + (\mathrm{dispersion})^{2}},
\end{equation}
where  `dispersion' corresponds to 28.4 m\AA/pixel, 31 m\AA/pixel and 227.3 m\AA/pixel for the three channels respectively \citep{Singh2019}. The pixel size of 6.5 $\mu$m for visible spectral channels and 25 $\mu$m for IR channel result in pixel scales as 1.25 arcsec pixel$^{-1}$ and 4.8 arcsec pixel$^{-1}$ respectively. It should be realised from the optical layout of VELC \citep[Figure 1 in][]{kumar2018optical}, the set of 4 slits in the slit plane placed in the optical path is incident on the spectrograph. We vary slit-width as 20 $\mu$m, 40 $\mu$m and 60 $\mu$m and investigate their effects on the instrument output (see Section \ref{sec:results}). The density and emission measure for the different scenarios are supplied as inputs to the CHIANTI. The synthetic {spectra} is computed using IDL procedures of CHIANTI, {\it ch\_synthetic.pro} and {\it make\_chianti\_spec.pro}. {The instrumental FWHM calculated using Equation \ref{eq:fwhm} is provided as an input to {\it make\_chianti\_spec.pro} for the instrument induced FWHM to the synthetic line.} The peak of the intensity obtained {having the physical unit of photons cm$^{-2}$ sr$^{-1}$ s$^{-1}$  \AA$^{-1}$ is then converted to equivalent photoelectrons incident on each pixel per second using the following relation},
\begin{equation}
\label{eq-peakPhot}
    \mathrm{ph\_elecs_{peak}} = \mathrm{(peak \; intensity) \times area \times (solid \; angle) \times dispersion \times efficiency} {\;} \mathrm{ph.electrons/pixel/s},
\end{equation}
where the area of the VELC primary is obtained taking the diameter of the primary mirror as 195 mm \citep{kumar2018optical}. The solid angle subtended is calculated taking the {slit scale} and pixel scale along the horizontal and vertical dimensions respectively. The efficiency of the spectral channels in visible and IR wavelengths is taken to be $\sim$5 \% and $\sim$4 \% respectively \citep{Singh2019}. 


The instrument contribution to the scattered intensity and noise arising due to the detector is also included in the simulated spectra. Scatter studies for the continuum channel of VELC (observing at 5000 \AA\ with 10 \AA\ pass-band) was done using  Advanced System Analysis Program (ASAP) by \citet{Venkata17}. {As VELC has narrow band filters for the three channels, we scaled this scattered intensity values taking into account the pass-band of individual channels.}  The scatter of the continuum was scaled to the spectral channels using, 

\begin{equation}
\label{eq-scatter}
    \mathrm{scatter_{spec}} = \frac{\mathrm{dispersion}}{10}\times\frac{\mathrm{slit \; scale \times (pixel \; scale) _{spectral}}}{\mathrm{(pixel \; scale)^{2} _{continuum}}} \times \mathrm{(Scatter_{continuum})} {\;} \mathrm{ph.electrons/pixel/s},
\end{equation}
which gives the number of scattered {photoelectrons} in the spectral channel generated in a pixel every second. The {pixel scale} for the continuum channel is 2.5 arcsec pixel$^{-1}$. The constant factor in the Equation \ref{eq-scatter} is obtained using the procedure for conversion followed in \citet{Patel2018}. The scatter is assumed to be circularly symmetric and is added to the synthetic spectra obtained using CHIANTI in Equation \ref{eq-peakPhot}. { This could be considered as the worst case scenario as scatter is inversely related to the square of incident wavelength.} The final spectra obtained is then used for the further analysis.
\\

\section{Analysis and Results}
    \label{sec:results}
The synthetic spectra for each channel is added with the dark noise associated with individual detectors. For the visible spectral channels using CMOS detector the dark noise ({\it D}) is up to 15 electrons where as in the high gain mode of the IR CCD, it is 42 electrons with readout noise ({\it R}) of 2 and 80 electrons respectively \citep[VELC team]{Singh2019} . {The photon noise ({\it p}) is calculated as the square root of the total { photoelectrons} generated including the coronal signal photons obtained in the spectra and the scattered photons.} The resulting signal to noise ratio (SNR) is then calculated as,

\begin{equation} 
\label{SNR}
\mathrm{SNR} = \frac{S}{\sqrt{p^2 + D^2 + R^2}},
\end{equation}
where S is the number of incident signal { photoelectrons} which in the cases analysed will be the photons determined from synthetic spectral signal. The SNR is calculated for the peak intensity of the simulated spectra and also at $\pm$0.5 \AA\ from the peak intensity wavelength. It is required to have an idea about the signal strength at $\pm$0.5 \AA\ as it will be helpful to determine sufficient signal requirements near the wings of the spectral line for obtaining a better fit.
\\

\subsection{Slit-width Optimization}
\label{sec-swopt}

It was reported by \citet{Singh2019} that the slit-width of VELC needs to be increased for to achieve the desired science goals. For obtaining an optimised value of slit-width we used the synthetic spectra for analysis. The synthetic spectra was generated for a quiet-Sun condition considering an average coronal temperature of 10$^{6.25}$ K for the heights ranging from 1.1 R$_\odot$ to 1.5 R$_\odot$ in steps of 0.1 R$_\odot$. The emission measure and electron density for simulation using the coronal parameters \citep{Baumbach37, allen1973book} at these five heights are:
\begin{itemize}
    \item log(EM) = [27, 26.3, 25.8, 25.36, 25]
    \item log(n$_e$) = [8.2, 7.85, 7.6, 7.38, 7.2].
\end{itemize} 
The spectra were generated for the three channels at the five heights including the scattered intensity and noises. The SNR per pixel per second at the peak intensity and at $\pm$0.5 \AA\ from the peak intensity wavelength were calculated for different slit-widths are tabulated in Table \ref{tab:diffTemp}.

\begin{table}[!ht]
    \centering
    \begin{tabular}{cccccccc}
    \hline \\

& Distance (R$_\odot$) & 1.1 & 1.2 & 1.3 & 1.4 & 1.5 \\ \hline \\
 \multicolumn{7}{c}{Slit width = 20 $\mu$m} \\ \hline \\    
          
\multirow{4}{*}{5303 \AA} & Peak & 659 & 182 & 80 & 41 & 25 \\
& Peak + scatter & 720 & 223 & 110 & 65 & 43 \\
& SNR (peak) & 21.39 & 8.56 & 4.34 & 2.39 & 1.512 \\
& SNR ($\pm$0.5 \AA) & 6.71 & 2.22 & 1.02 & 0.56 & 0.37 \\ \hline \\

\multirow{4}{*}{7892 \AA} & Peak &  53 & 17 &  9 &  5 & 3 \\
& Peak + scatter & 120 & 62 & 42 & 31 & 23 \\
& SNR (peak) &  3.15 & 1.08 & 0.58 & 0.33 &  0.19 \\
& SNR ($\pm$0.5 \AA) & 1.5 & 0.53 & 0.31 & 0.18 & 0.13 \\ \hline \\

\multirow{4}{*}{10747 \AA} & Peak & 14349 & 5439 & 2820  & 1619 & 1041 \\
& Peak + scatter & 16212 & 6681 & 3731 & 2344 & 1593 \\
& SNR (peak) & 113.02 & 64.06 & 41.63 & 27.82 & 19.64 \\
& SNR ($\pm$0.5 \AA) & 95.48 & 52.01 & 33.02 & 21.68 & 15.31 \\ \hline \\

 \multicolumn{7}{c}{Slit width = 40 $\mu$m} \\ \hline \\ 
\multirow{4}{*}{5303 \AA} & Peak & 1287 & 355 & 155 & 80 & 48 \\
& Peak + scatter & 1409 & 436 & 215 & 128 & 84 \\
& SNR (peak) & 31.79 & 13.76 & 7.35 & 4.23 & 2.71 \\
& SNR ($\pm$0.5 \AA) & 11.5 & 4.09 & 1.95 & 1.05 & 0.66 \\ \hline \\

\multirow{4}{*}{7892 \AA} & Peak & 104 & 34 & 17 & 9 & 6 \\
& Peak + scatter & 237 & 123 & 82 & 61 & 46 \\
& SNR (peak) & 5.69 & 2.09 & 1.08 & 0.58 & 0.39 \\
& SNR ($\pm$0.5 \AA) & 2.65 &  0.98 & 0.51 & 0.29 & 0.18 \\ \hline \\

\multirow{4}{*}{10747 \AA} & Peak & 28018 & 10620 & 5506 & 3161 & 2033 \\
& Peak + scatter & 31744 & 13104 & 7328 & 4610 & 3137 \\
& SNR (peak) & 162.34 & 95.41 & 64.55 & 45.02 & 32.97 \\
& SNR ($\pm$0.5 \AA) & 137.81 & 77.52 & 50.88 & 34.54 & 25.15 \\ \hline \\

 \multicolumn{7}{c}{Slit width = 60 $\mu$m} \\ \hline \\ 
\multirow{4}{*}{5303 \AA} & Peak & 1861 & 513 & 224 & 115 & 69 \\
& Peak + scatter & 2043 & 635 & 313 & 186 & 123 \\
& SNR (peak) & 39.04 & 17.45 & 9.62 & 5.64 & 3.67 \\
& SNR ($\pm$0.5 \AA) & 15.76 & 6.28 & 2.94 & 1.59 & 1.03 \\ \hline \\

\multirow{4}{*}{7892 \AA} & Peak & 152 & 50 & 24 & 14 & 9 \\
& Peak + scatter & 351 & 183 & 121 & 92 & 68 \\
& SNR (peak) & 7.78 & 2.99 & 1.51 & 0.89 & 0.58 \\
& SNR ($\pm$0.5 \AA) & 3.63 & 1.37 & 0.71 & 0.39 & 0.29 \\ \hline \\

\multirow{4}{*}{10747 \AA} & Peak & 40476 & 15342 & 7954 & 4567 & 2937 \\
& Peak + scatter & 46065 & 19068 & 10686 & 6741 & 4593 \\
& SNR (peak) & 196.93 & 117.28 & 80.66 & 57.37 & 42.81 \\
& SNR ($\pm$0.5 \AA) & 168.1 & 95.62 & 63.60 & 43.80 & 32.37 \\ \hline
    \end{tabular}
    \caption{Optimization of the slit-width using the SNR calculations for slit-widths of 20 $\mu$m, 40, $\mu$m and 60 $\mu$m for log(T) = 6.25. The parameters peak and peak+scatter are in the unit of {photoelectrons/pixel/second.}}
    \label{tab:diffTemp}
\end{table}

It can be seen from Table \ref{tab:diffTemp} that as the slit-width is increased, there is an improvement in the SNR of all the channels. The increased SNR is also observed at the wings of the spectral line used for analysis. However, it should be noted that there has also been increase in the number of photons incident on the detector. Increasing the number of incident photons also impose the challenge of attaining sufficient SNR without saturating the detector. The CMOS sensors used for visible spectral channels have full well capacity of $\sim$30000 electrons in both high and low gain where as INGAS (InGaS) used for IR detector it is $\sim$30300 electrons in high-gain mode \citep{Singh2019}. The IR channel is also equipped with spectro-polarimeter mode which will operate in high-gain mode with a fixed exposure time of 500 ms. When operated in this mode, it can be noted from  Table \ref{tab:diffTemp} that for a slit-width of 60 $\mu$m, the incident photons count to $\sim$23033 electrons at 1.1 R$_\odot$ which is $\sim$77\% of the full well capacity. Thus, the slit-width needs to be $\leq$60$\mu$m. Therefore, we tested the SNR with 50 $\mu$m slit-width keeping rest of the parameters same as above. The reults are tabulated in Table \ref{tab:50mic}.

\begin{table}[]
    \centering
    \begin{tabular}{ccccccc}
     \hline \\
\multicolumn{7}{c}{Slit width = 50 $\mu$m} \\ \hline \\ 
& Distance (R$_\odot$) & 1.1 & 1.2 & 1.3 & 1.4 & 1.5 \\ \hline \\
\multirow{4}{*}{5303 \AA} & Peak & 1582 & 437 & 190 & 98 & 59 \\
& Peak + scatter & 1734 & 539 & 265 & 157 & 104 \\
& SNR (peak) & 35.71 & 15.76 & 8.55 & 4.98 & 3.23 \\
& SNR ($\pm$0.5 \AA) & 13.64 & 4.99 & 2.46 & 1.35 & 0.82 \\ \hline \\

\multirow{4}{*}{7892 \AA} & Peak & 128 & 42 & 20 & 12 & 7 \\
& Peak + scatter & 294 & 153 & 101 & 77 & 56 \\
& SNR (peak) & 6.77 & 2.55 & 1.26 & 0.77 & 0.45 \\
& SNR ($\pm$0.5 \AA) & 3.16 & 1.15 & 0.61 & 0.34 & 0.23 \\ \hline \\

\multirow{4}{*}{10747 \AA} & Peak & 34423 & 13047 & 6765 & 3884 & 2498 \\
& Peak + scatter & 39080 & 16152 & 9042 & 5695 & 3878 \\
& SNR (peak) & 180.95 & 107.19 & 73.23 & 51.66 & 38.24 \\
& SNR ($\pm$0.5 \AA) & 154.01 & 87.22 & 57.71 & 39.51 & 29.02 \\ \hline
    \end{tabular}
    \caption{SNR for the three spectral channels of VELC for optimized slit-width of 50 $\mu$m for log(T) = 6.25.}
    \label{tab:50mic}
\end{table}

It could also be noticed from Table \ref{tab:50mic}:
\begin{enumerate}[(i)]
    \item  With a slit-width of 50 $\mu$m, the IR spectro-polarimeter mode could get an incident photon count of 19540 at 1.1 R$_\odot$ having sufficient SNR at the same time.
    \item The slit-width of 50 $\mu$m leads to a photon count contributing to $\sim$65\% of the full well capacity for IR channel at 1.1 R$_\odot$ in spectro-polarimeter mode. This is a sufficient margin to account for the flaring conditions or intensity enhancement in the coronal structures.
    \item  The SNR at the peak and wings decrease with height for all the channels. This implies that subsequent frames may be added post-facto to enhance the SNR.
    \item  SNR for Fe XI (7892 \AA) channel {is lower than the other two channels} for the selected simulation parameters. \\
\end{enumerate}


\subsection{Effect of Temperature on SNR}

As the corona contains plasma of different temperatures, after optimizing the slit-width to 50 $\mu$m, we analysed the effect of different plasma temperatures on the performance of the spectral channels. The spectra are synthesized for three channels at a height of 1.1 R$_\odot$ taking electron density of 10$^{8.2}$ cm$^{-3}$ and EM = 10$^{27}$ cm$^{-5}$. The temperature is varied from log(T) = 6.0 to log(T) = 6.5 in steps of 0.1. The scattered intensity and noise introduced by the instrument is also added accordingly. The result is tabulated in Table \ref{tab:50Temp}. It is noticed that the three channels show maximum SNR corresponding to different temperatures. This directly implies the importance of these lines for temperature diagnostics of the corona. It could be noted that 7892 \AA\ channel shows good SNR for relatively cool plasma as compared to other channels as the corresponding Fe XI ion formation temperature is comparatively lower than the other two ions in consideration. Looking at the values of SNRs for the three channels in Table \ref{tab:50Temp} also reveals that if the study using these three lines are combined then it will be helpful to investigate plasma over a wide range of temperatures.

\begin{table}[]
    \centering
    \begin{tabular}{cccccccccc}
     \hline \\
\multicolumn{7}{c}{Slit width = 50 $\mu$m} \\ \hline \\ 
& log(T) & 6.0 & 6.1 & 6.2 & 6.3 & 6.4 & 6.5 \\ \hline \\
\multirow{4}{*}{5303 \AA} & Peak & 11 & 249 & 1237 & 1391 & 390 & 68 \\
& Peak + scatter & 163 & 401 & 1389 & 1543 & 542 & 220 \\
& SNR (peak) & 0.55 & 9.92 & 30.75 & 33.04 & 14.04 & 3.21 \\
& SNR ($\pm$0.5 \AA) & 0.05 & 1.76 & 9.98 & 13.81 & 6.02 & 1.48 \\ \hline \\

\multirow{4}{*}{7892 \AA} & Peak & 1025 & 1043 & 342 & 36 & 2 & 1 \\
& Peak + scatter & 1191 & 1209 & 508 & 202 & 168 & 167 \\
& SNR (peak) & 28.95 & 29.24 & 14.31 & 2.21 & 0.13 & 0.06 \\
& SNR ($\pm$0.5 \AA) & 12.63 & 14.97 & 7.04 & 0.98 & 0.05 & 0.05 \\ \hline \\

\multirow{4}{*}{10747 \AA} & Peak & 3183 & 23479 & 42559 & 19635 & 2495 & 214 \\
& Peak + scatter & 7840 & 28136 & 47216 & 24292 & 7152 & 4871 \\
& SNR (peak) & 45.23 & 147.76 & 202.14 & 134.21 & 38.21 & 4.81 \\
& SNR ($\pm$0.5 \AA) & 25.65 & 117.75 & 172.51 & 110.26 & 23.67 & 2.37 \\ \hline
    \end{tabular}
    \caption{Effect of different plasma temperatures on the SNRs of VELC spectral channels estimated at 1.1~R$_\odot$.}
    \label{tab:50Temp}
\end{table}

\begin{figure}[!ht]
    \centering
    \centerline{\hspace*{0.05\textwidth}
               \includegraphics[width=0.33\textwidth,clip=]{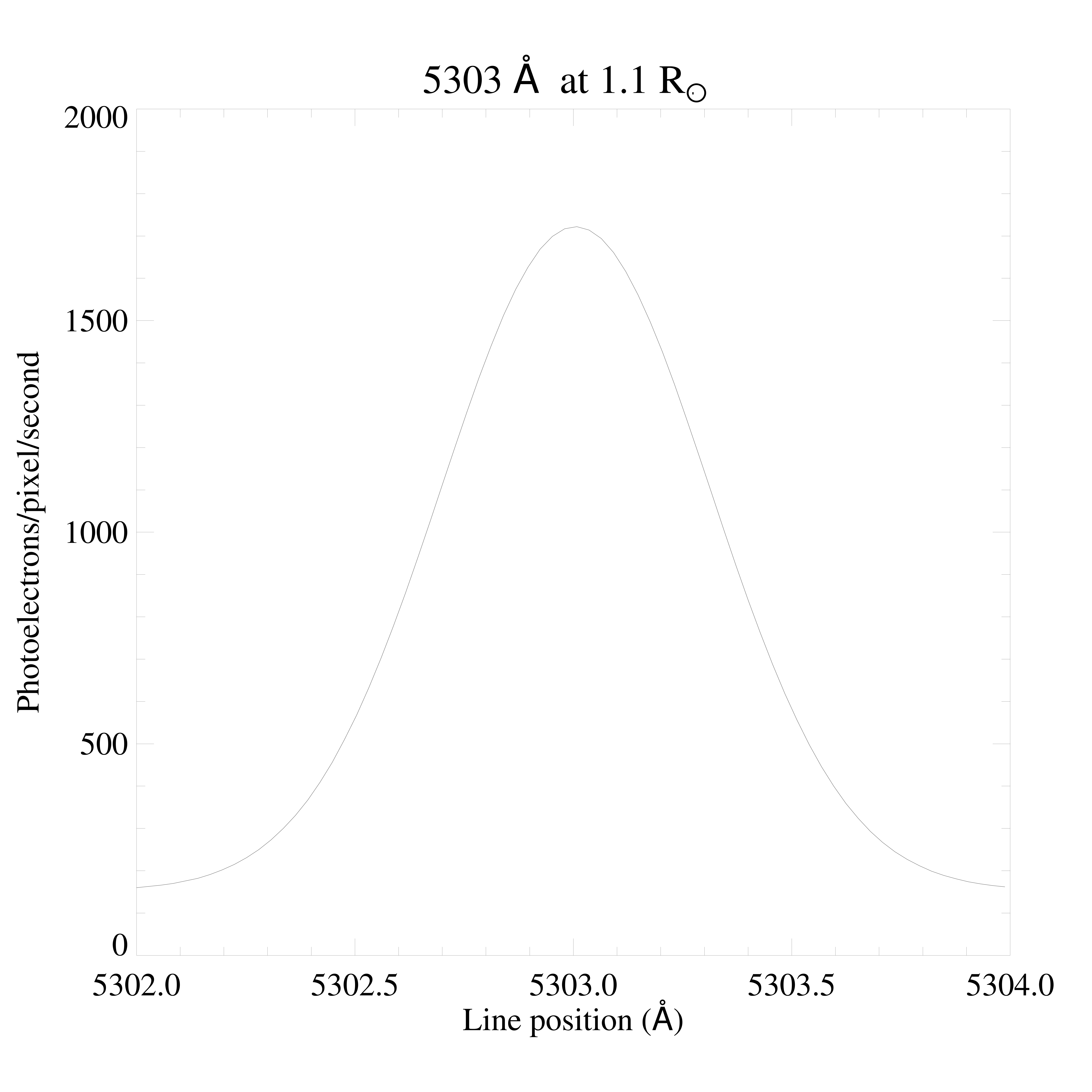}
               \hspace*{0.002\textwidth}
               \includegraphics[width=0.33\textwidth,clip=]{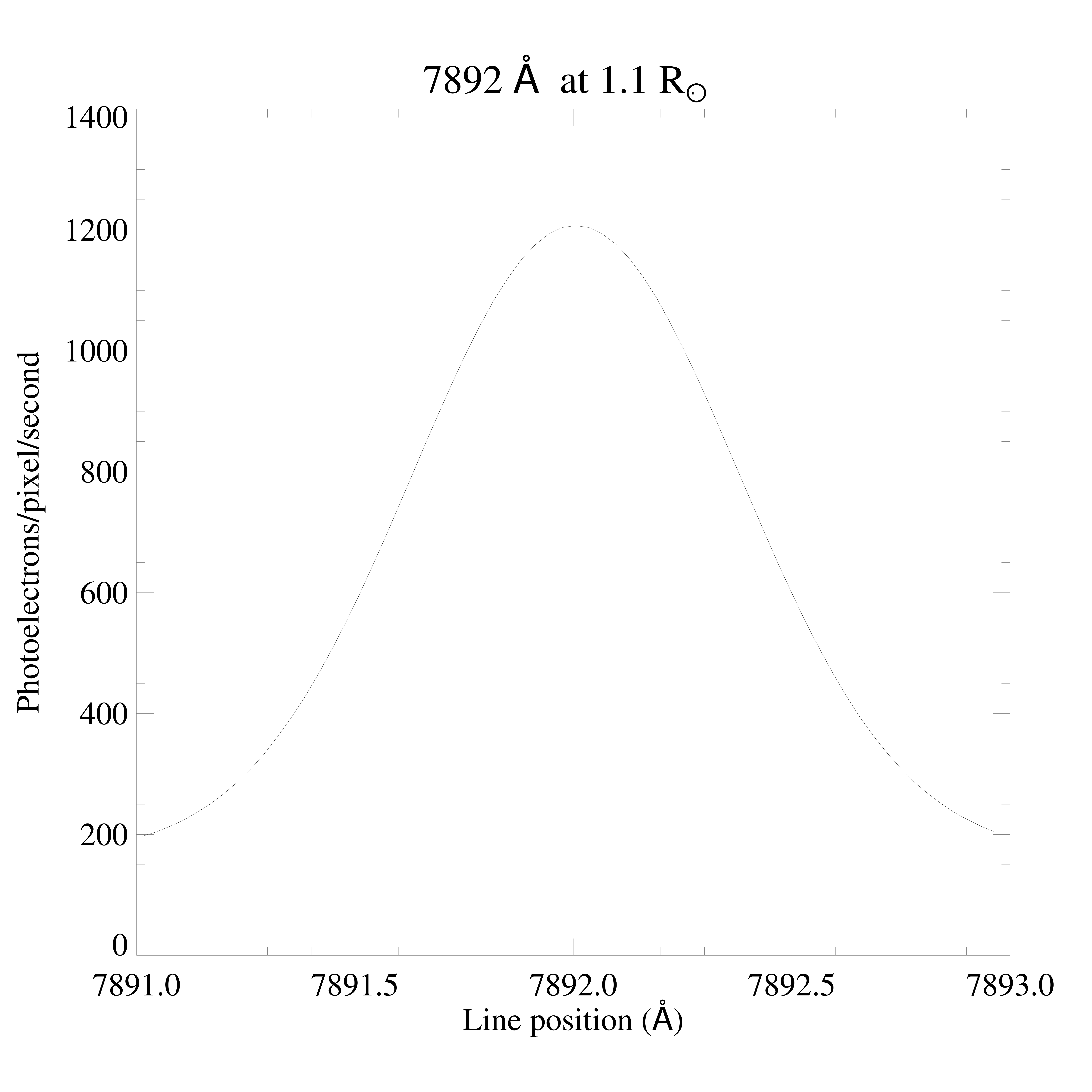}
               \includegraphics[width=0.33\textwidth,clip=]{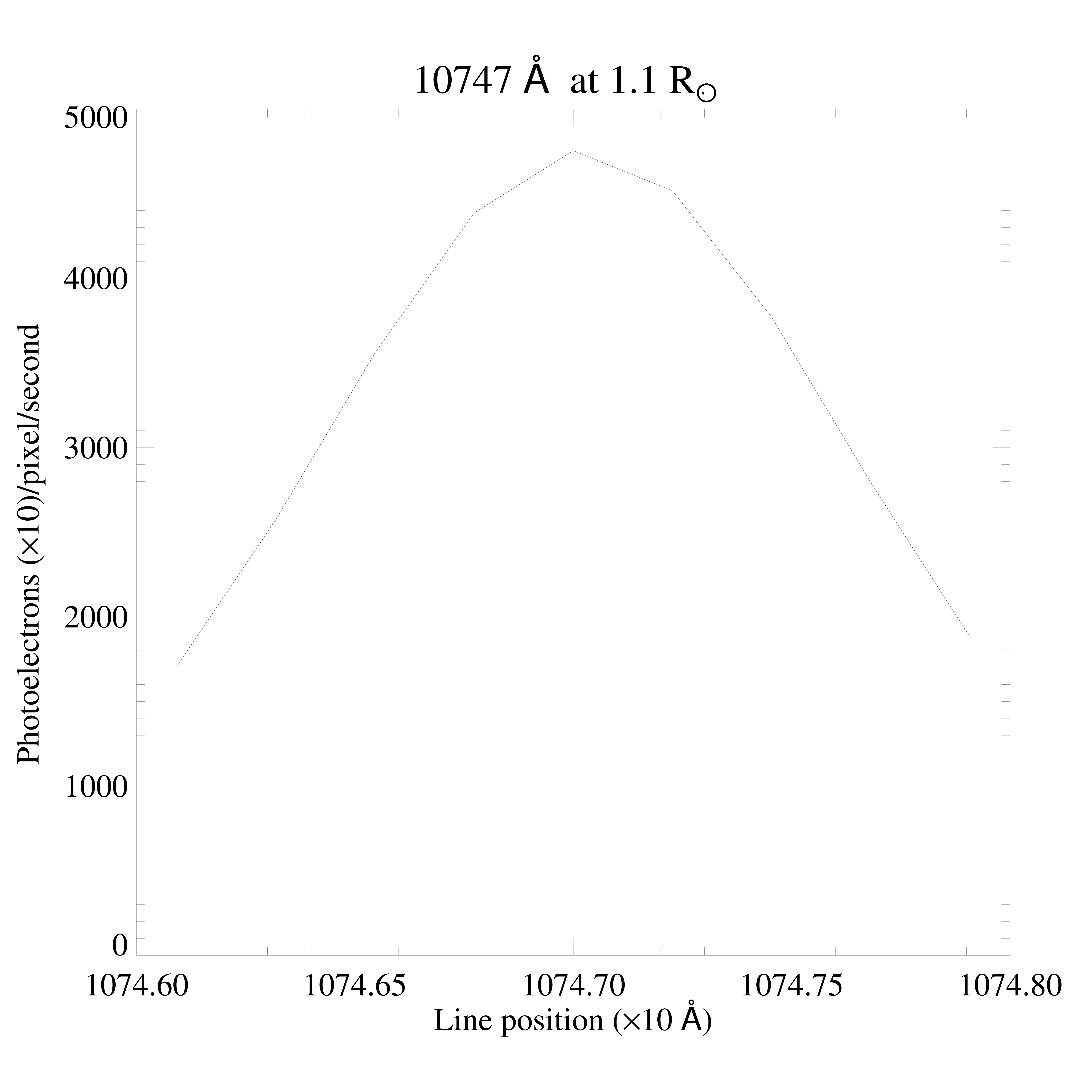}
              }
    \centerline{    
      \hspace{0.175\textwidth}  \color{black}{(a)}
      \hspace{0.3\textwidth}  \color{black}{(b)}
      \hspace{0.3\textwidth}  \color{black}{(c)}
         \hfill}
    \caption{Synthetic spectra for VELC at 1.1 R$_\odot$ for (a) 5303 \AA, (b) 7892 \AA, and (c) 10747 \AA\ for 50 $\mu$m slit-width at their respective line formation temperature. {Photons due to instrument scattering are added to the spectra.}}
    \label{fig:50mic_peak}
\end{figure}

We then synthesized the lines at their respective peak formation temperature and proceeded as above. 
The temperatures chosen are close to the peak line formation temperature for these lines which are 10$^{6.27}$~K, 10$^{6.1}$ K, and 10$^{6.2}$ K for 5303 \AA, 7892 \AA, and 10747 \AA\ respectively \citep{allen1973book}. We synthesized these three spectral lines at heights ranging from 1.1 to 1.5 R$_\odot$ with emission measure and electron density as mentioned in Section \ref{sec-swopt} for slit-width of 50 $\mu$m considering the scatter and noise addition as for previous cases. Figure \ref{fig:50mic_peak} shows such synthetic spectra expected to be observed by VELC at 1.1 R$_\odot$. The spectra also includes added scatter and noise values from the instrument at this height. The results of this analysis for the mentioned heights is summarised in Table \ref{tab:50micpeak} which reveals peak SNR at the line peak as well as at the wings. On comparison with Table \ref{tab:50mic}, it could be seen that the observed line intensity will have sufficient SNR up to larger distances for line emission corresponding to their peak temperatures for all the spectral channels. For the cases of 5303 \AA\ and 7892 \AA\ when the SNR becomes $\leq$ 5, then it could be enhanced by pixel or frame binning as required. \\

\begin{table}[]
    \centering
    \begin{tabular}{ccccccc}
     \hline \\
\multicolumn{7}{c}{Slit width = 50 $\mu$m} \\ \hline \\ 
& Distance (R$_\odot$) & 1.1 & 1.2 & 1.3 & 1.4 & 1.5 \\ \hline \\
\multirow{4}{*}{\begin{tabular}[c]{@{}c@{}}5303 \AA\\[1ex] log(T) = 6.27 \end{tabular}} & 
 Peak & 1611 & 486 & 209 & 110 & 65 \\
& Peak + scatter & 1763 & 570 & 284 & 169 & 110 \\
& SNR (peak) & 36.1 & 17.19 & 9.22 & 5.51 & 3.53 \\
& SNR ($\pm$0.5 \AA) & 13.64 & 4.99 & 2.46 & 1.35 & 0.82 \\ \hline \\


\multirow{4}{*}{\begin{tabular}[c]{@{}c@{}}7892 \AA\\[1ex] log(T) = 6.1 \end{tabular}} & Peak & 1043 & 334 & 161 & 88 & 56  \\
& Peak + scatter & 1209 & 445 & 242 & 153 & 105 \\
& SNR (peak) & 29.24 & 14.07 & 8.15 & 4.94 & 3.31 \\
& SNR ($\pm$0.5 \AA) & 14.97 & 6.31 & 3.40 & 2.03 & 1.32 \\ \hline \\

\multirow{4}{*}{\begin{tabular}[c]{@{}c@{}}10747 \AA\\[1ex] log(T) = 6.2 \end{tabular}} & Peak & 42559 & 16095 & 8331 & 4777 & 3070 \\
& Peak + scatter & 47216 & 19200 & 10608 & 6588 & 4450 \\
& SNR (peak) & 202.14 & 120.42 & 82.90 & 59.04 & 44.13  \\
& SNR ($\pm$0.5 \AA) & 172.51 & 98.70 & 65.88 & 45.52 & 33.66  \\ \hline
    \end{tabular}
    \caption{SNR for VELC spectral channels for slit-width of 50 $\mu$m taking the line formation peak temperature for each channel.}
    \label{tab:50micpeak}
\end{table}

\begin{figure}[!ht]
    \centering
    \centerline{\hspace*{0.05\textwidth}
               \includegraphics[width=0.5\textwidth,clip=]{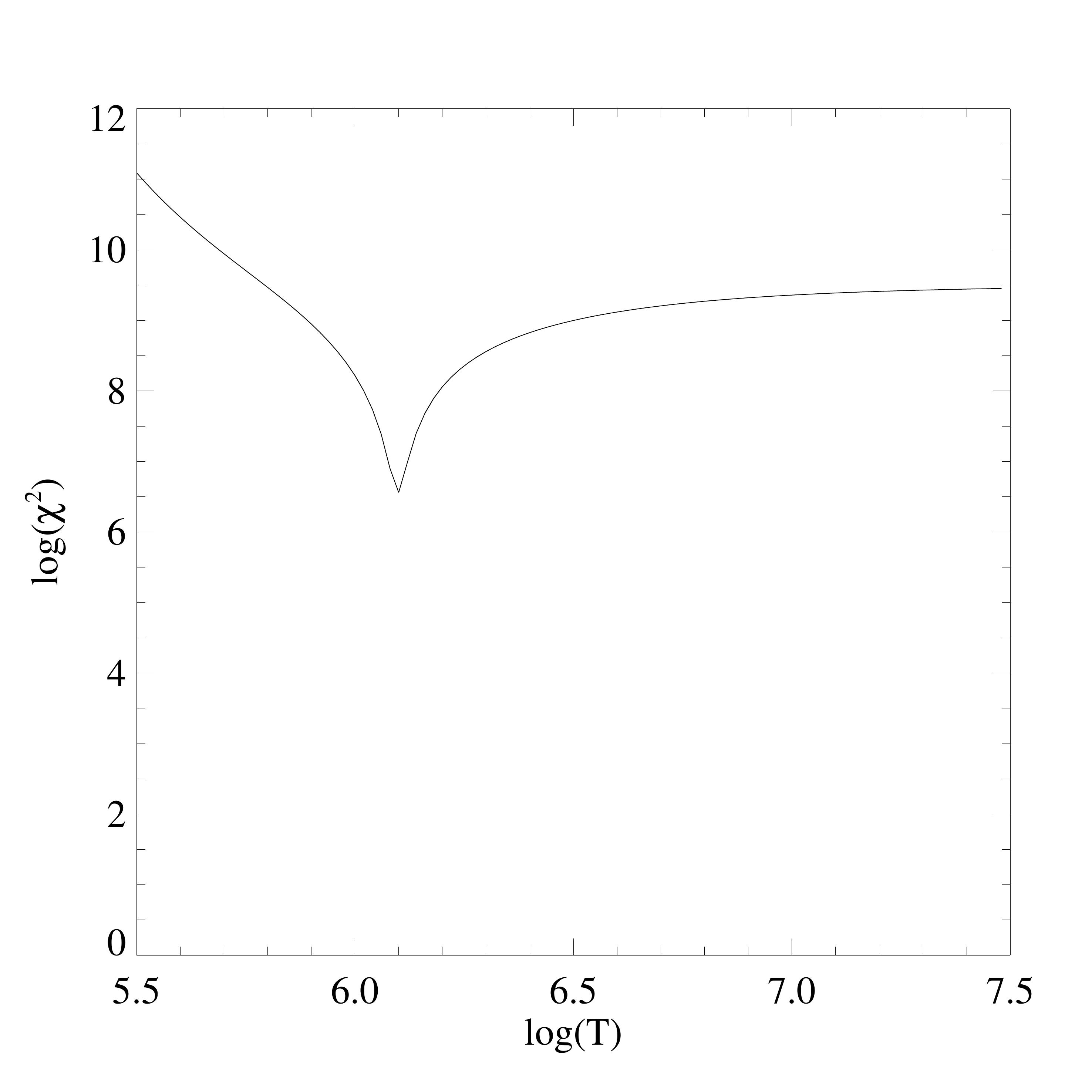}
               \hspace*{0.002\textwidth}
               \includegraphics[width=0.52\textwidth,clip=]{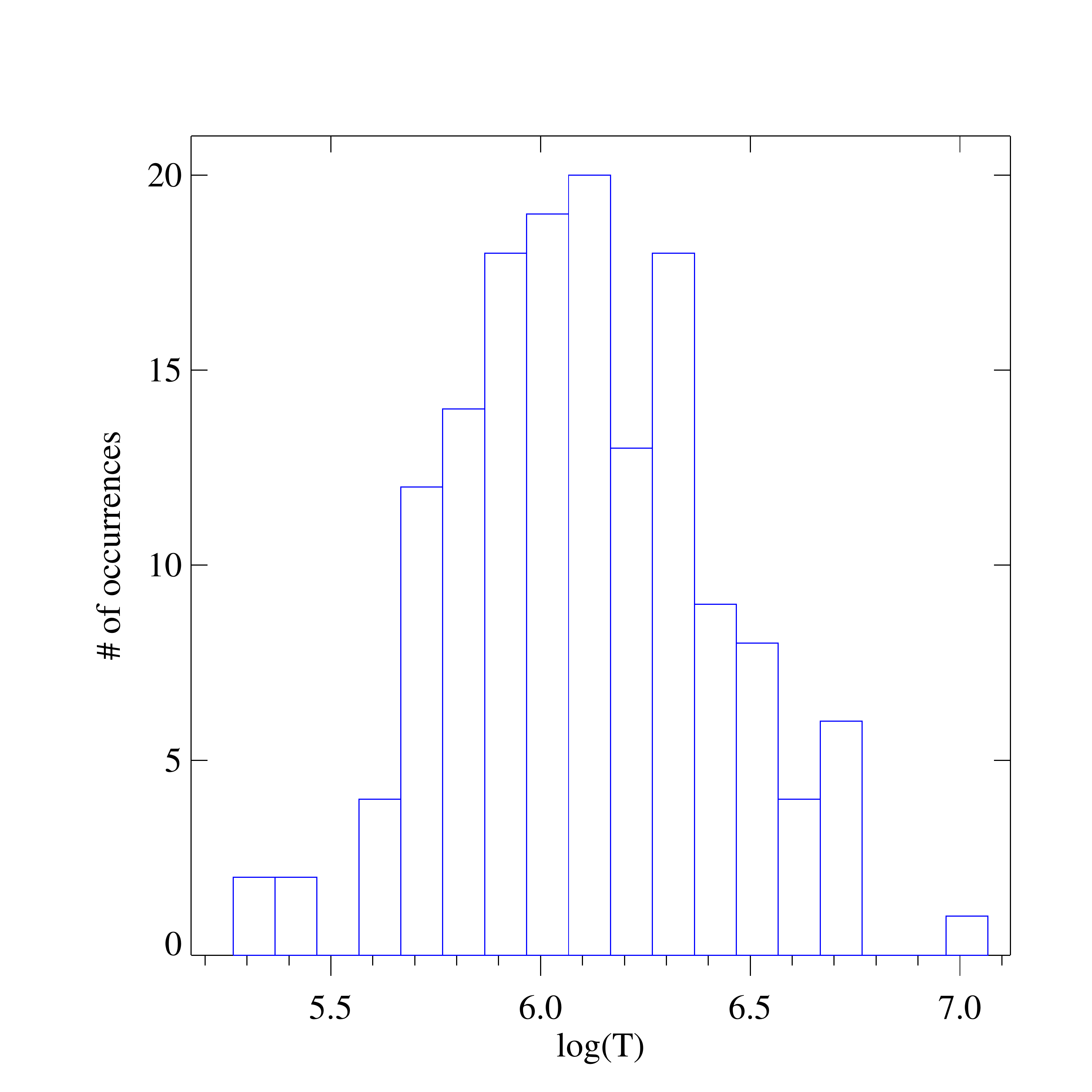}
              }
    \centerline{    
      \hspace{0.3\textwidth}  \color{black}{(a)}
      \hspace{0.4\textwidth}  \color{black}{(b)}
         \hfill}
    \caption{(a) Chi-square minimisation output as a function of temperature, (b) Temperature distribution with peak at log(T) = 6.1 used for MCMC simulation.}
    \label{fig:chi_comp}
\end{figure}

These cases have been simulated taking the isothermal corona with electron density calculated using the Baumbach model. As this coronal density model is based on white-light eclipse observations that include contributions from different temperatures, we used a temperature distribution to calculate the signal variation with height for the three channels. {We then compared} the densities obtained from Baumbach model with the values by the following relation:

\begin{equation}
\label{eq-scaleht}
    n_e = n_0e^{-(\frac{r-1.1}{H})},
\end{equation}
where n$_0$ is the electron density estimated at 1.1 R$_\odot$ using the {Baumbach model} and H is the scale height which is dependent on the temperature. Using the two estimates of the densities we performed a chi-square analysis to obtain the maximum match between the two models (Figure \ref{fig:chi_comp}a). We found that for log(T) = 6.1, density estimates using the two methods match well. Taking this as the peak value of temperature and width of 0.3 we generated 150 random numbers with a Gaussian distribution as shown in Figure \ref{fig:chi_comp}(b) to perform Markov chain Monte Carlo simulation \citep[MCMC;][]{Hastings1970} taking quiet-Sun densities and EM at the heights used previously with 50 $\mu$m slit-width. 

\begin{figure}[!ht]
    \centering
    \centerline{\hspace*{0.05\textwidth}
               \includegraphics[width=0.33\textwidth,clip=]{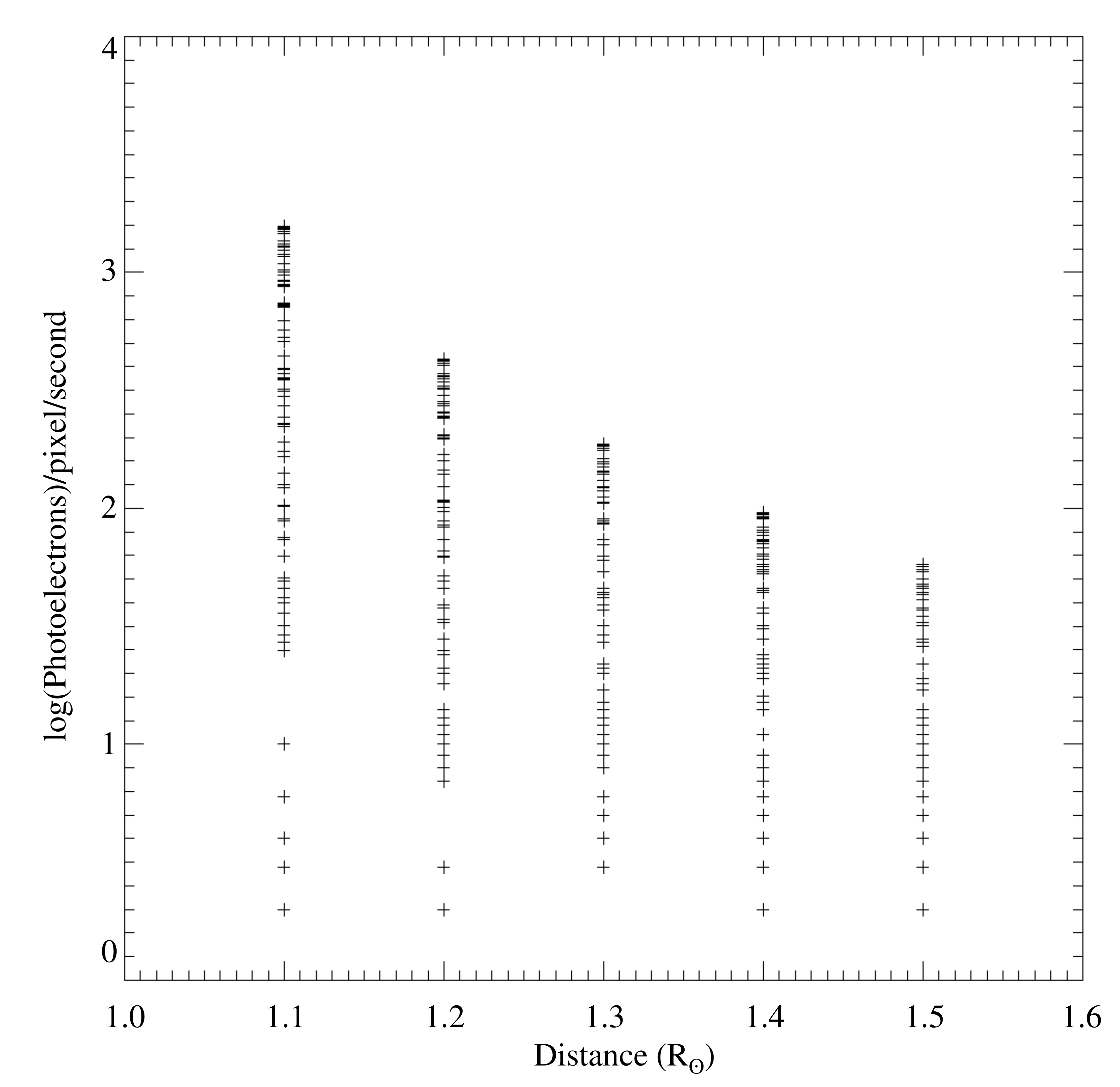}
               \hspace*{0.002\textwidth}
              \includegraphics[width=0.33\textwidth,clip=]{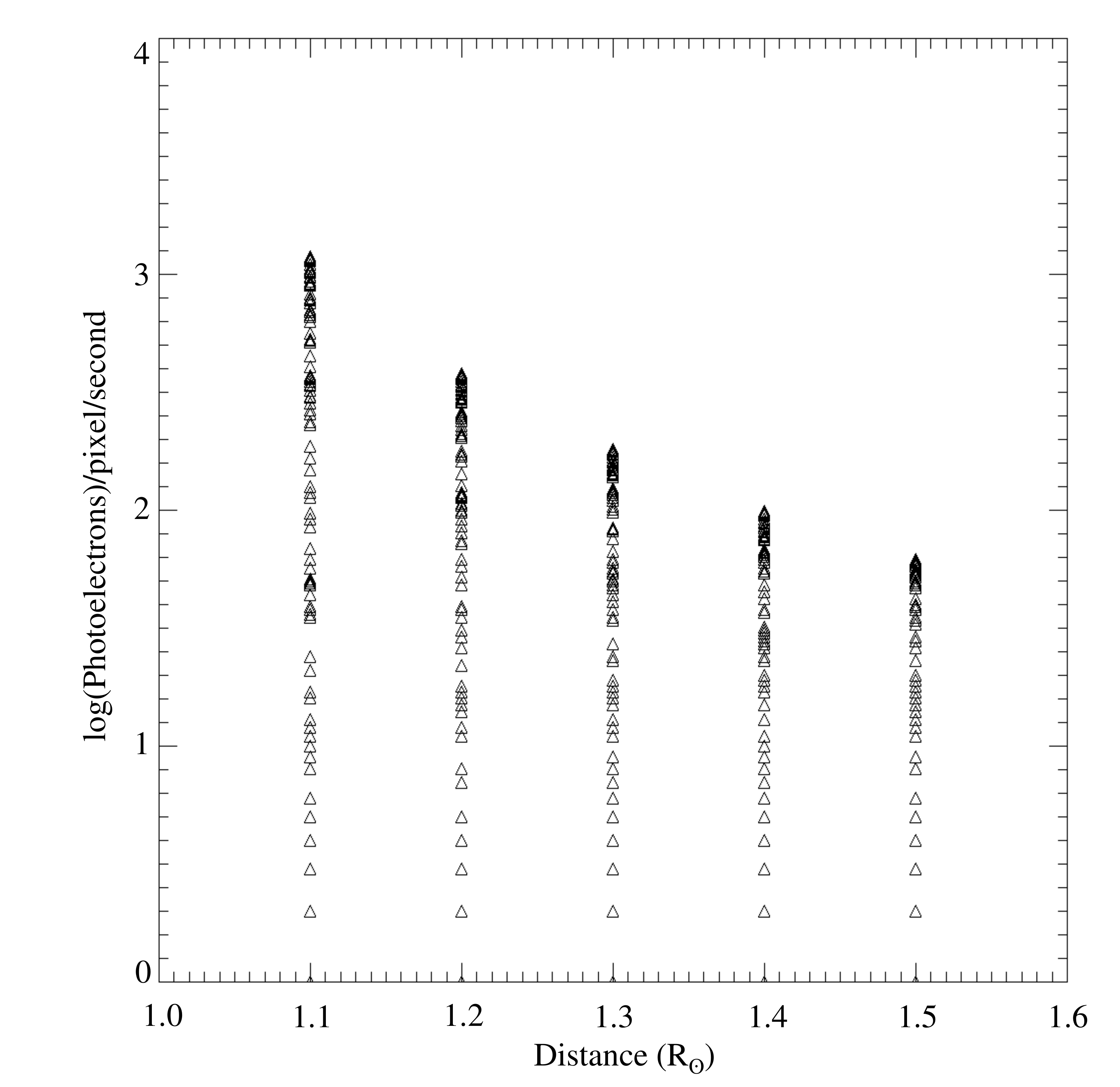}
              \includegraphics[width=0.33\textwidth,clip=]{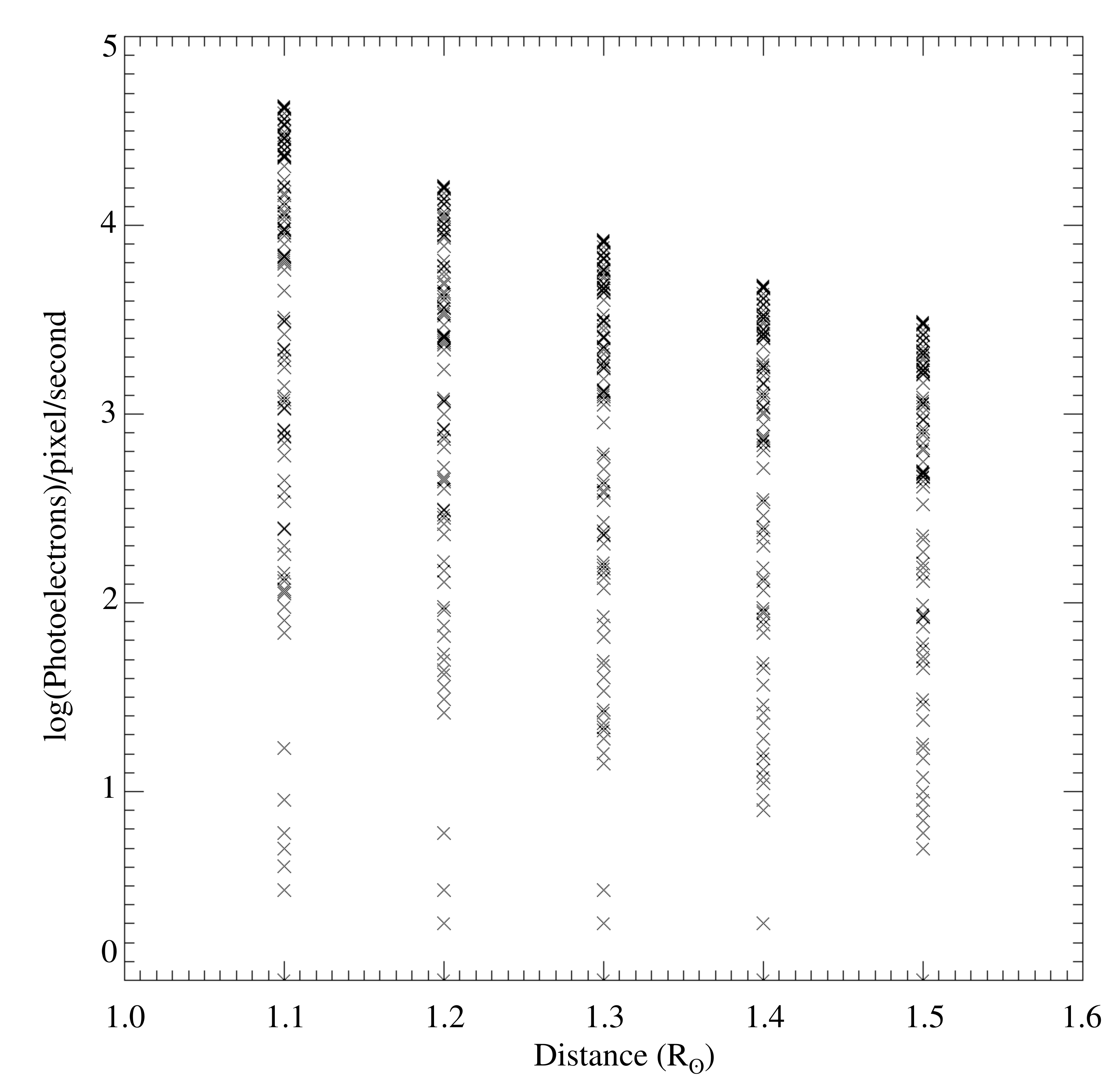}
              }
    \centerline{    
      \hspace{0.175\textwidth}  \color{black}{(a)}
      \hspace{0.3\textwidth}  \color{black}{(b)}
      \hspace{0.3\textwidth}  \color{black}{(c)}
         \hfill}
    \caption{Photoelectrons variation using synthetic spectra for VELC at different heights for (a) 5303 \AA, (b) 7892 \AA, and (c) 10747 \AA\ for 50 $\mu$m slit-width with temperature distribution. {It should be noted that the counts for IR channel is about an order of magnitude greater than the other two channels as the IR sensor pixels are larger than the visible sensor pixels.\\}}
    \label{fig:dist_height}
\end{figure}

We then estimated the photoelectrons generated in each pixel every second with temperature based on the distribution. The variation of the photoelectrons with height is shown in Figure \ref{fig:dist_height} for the three channels. It can be noted that there is a cluster of points near the top at each height for all channels. These values can be compared with the number of photoelectrons calculated at the line formation temperature for each channel as {specified} in Table \ref{tab:50micpeak}. Also when the points in the clusters in Figure \ref{fig:dist_height} are compared with the counts tabulated in {Table \ref{tab:50Temp}}, we could infer that the spread near the peak counts in each channel is due to the contribution from the temperature close to the peak line formation temperatures for the respective lines. Thus, even though there is a range of counts available from zero to thousands of photoelectrons each second, the maximum contribution will be observed due to the coronal structures contributing at and near the peak temperature for VELC channels.

\subsection{Effect of CME on Spectra}    
\begin{figure}
    \centering
    \centerline{\hspace*{0.05\textwidth}
               \includegraphics[width=0.5\textwidth,clip=]{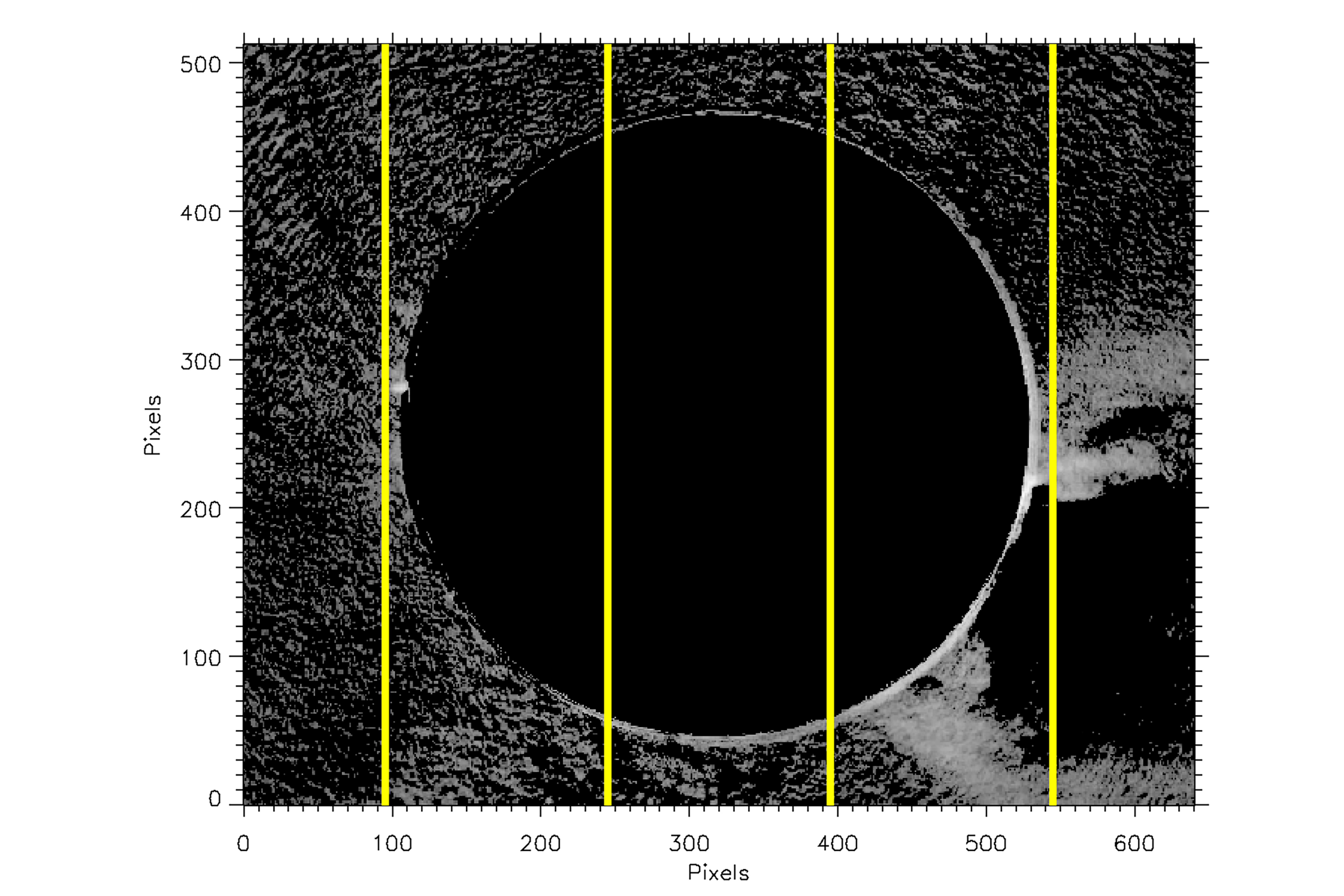}
               \hspace*{0.001\textwidth}
               \includegraphics[width=0.5\textwidth,clip=]{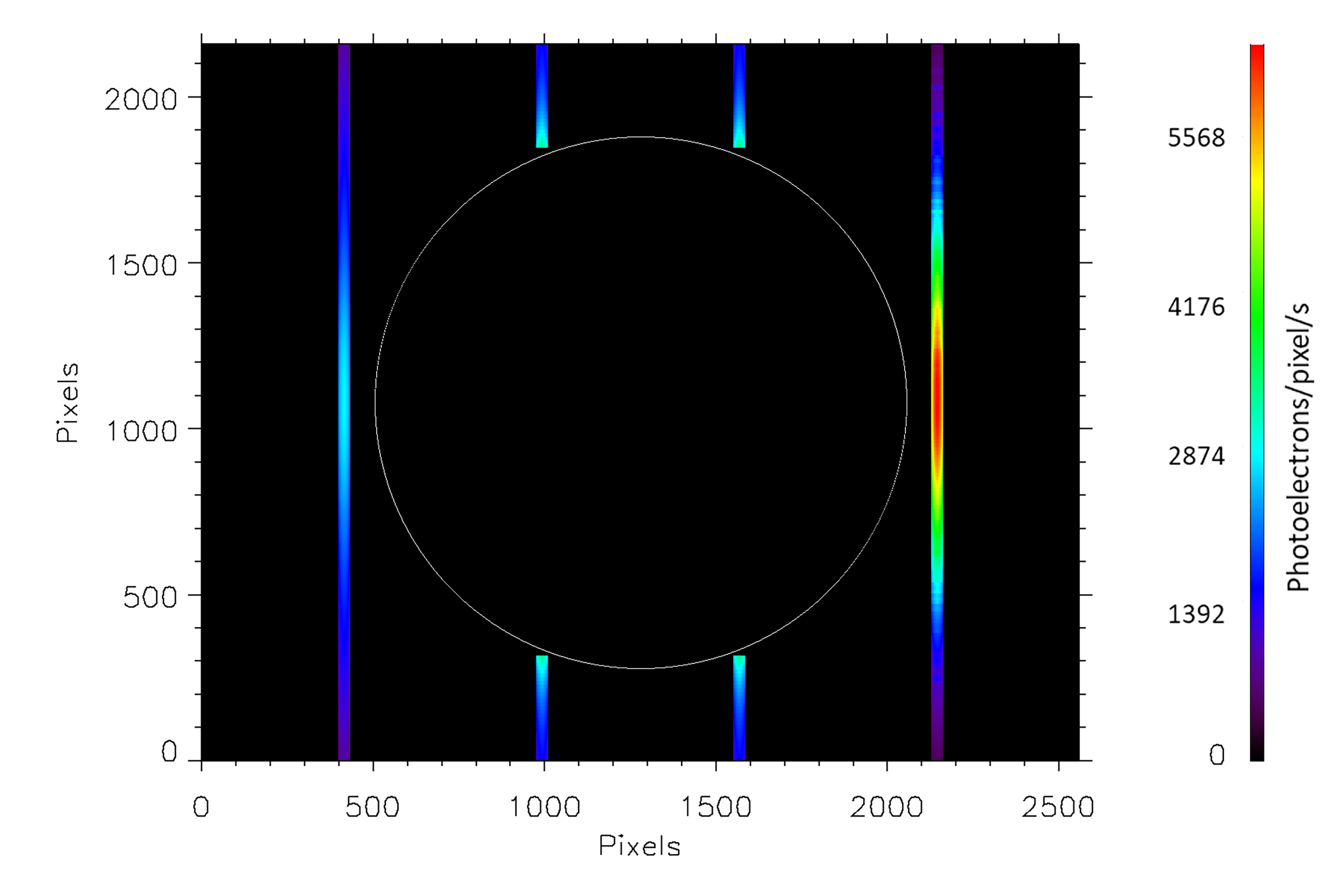}
              }
    \centerline{    
      \hspace{0.23\textwidth}  \color{black}{(a)}
      \hspace{0.5\textwidth}  \color{black}{(b)}
         \hfill}
      \vspace{0.01\textwidth}       
    \centerline{\hspace*{0.05\textwidth}
               \includegraphics[width=0.5\textwidth,clip=]{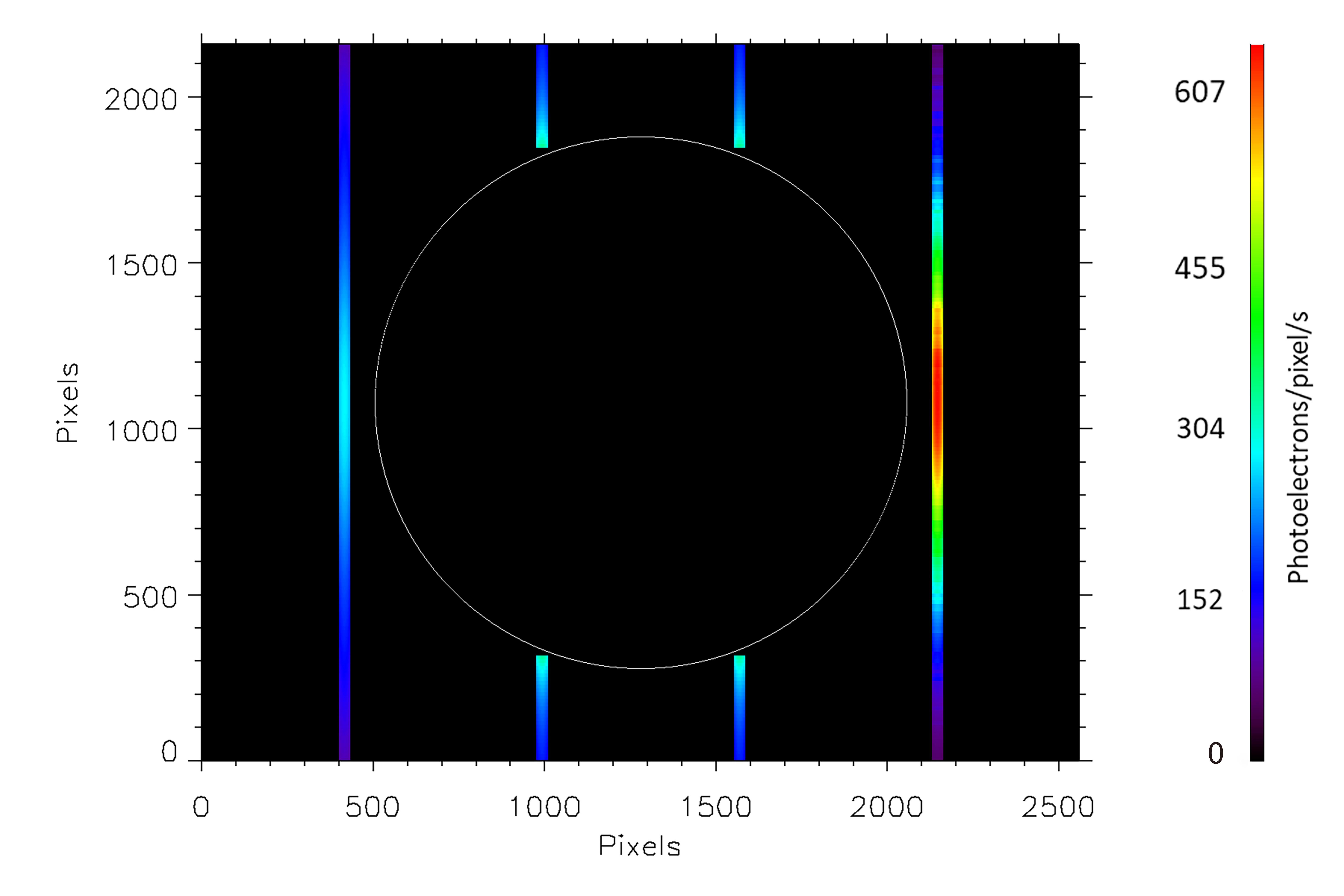}
               \hspace*{0.002\textwidth}
               \includegraphics[width=0.5\textwidth,clip=]{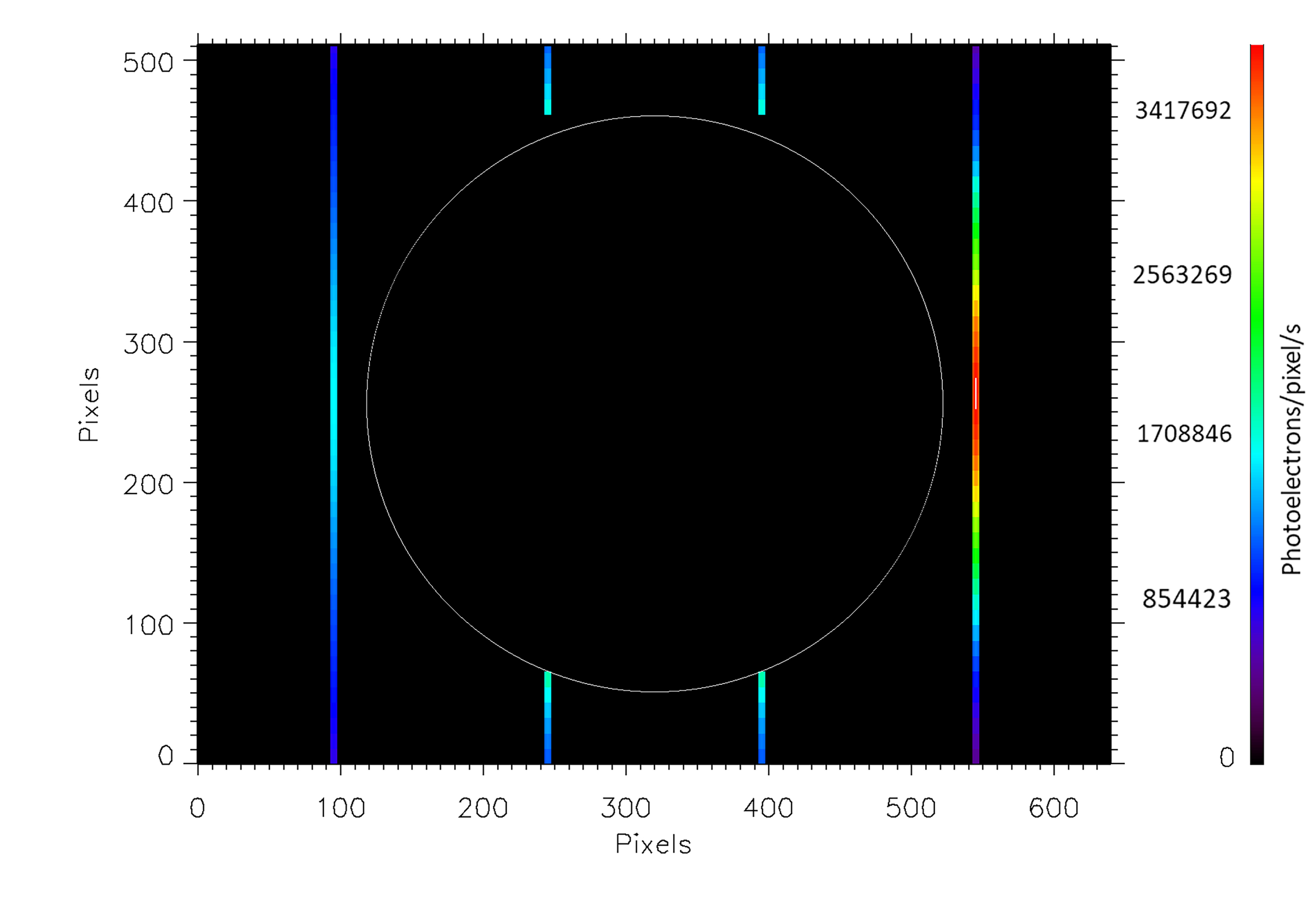}
              }
    \centerline{    
      \hspace{0.23\textwidth}  \color{black}{(c)}
      \hspace{0.5\textwidth}  \color{black}{(d)}
         \hfill}
    \caption{(a) VELC slit locations over-plotted in yellow on the KCor image of 2016-01-01 for VELC FOV in IR channel, (b), (c), and (d) Synthesized spectra for the four slits of VELC for 5303 \AA, 7892 \AA, and 10747 \AA\ respectively. It could be noticed that the region where CME is present, the intensity of the spectra is enhanced.}
    \label{fig:cmecase}
\end{figure}

We also analysed the performance of the instrument for a case when CME is passing through spectrograph slits with 50 $\mu$m slit-width. It should be noted that ground based coronagraph MLSO/KCor has FOV similar to VELC. We used a CME case that was observed by KCor on 2016-01-01. A reference image for the locations of slit in VELC FOV in spectral channels is shown in yellow superimposed on the KCor image as shown in Figure \ref{fig:cmecase}(a). Due to the difference in size of visible (2560 $\times$ 2160 pixels) and IR (640 $\times$ 512 pixels) channel detectors, the FOV covered is slightly different for the two. A circularly symmetric uniform coronal density based on Baumbach model \citep{Baumbach37} was used for synthesizing the spectra. Since the white light intensity is proportional to the electron density, we estimated the electron density by taking the ratio of CME and non-CME image. The non-CME image used here was an average intensity image of the images prior to CME occurrence.
The ratio provided an enhancement at the CME location with respect to the background corona. It was found that for this CME the maximum enhancement attained was $\sim$4 times above the background corona. This electron density was then used to synthesize the spectra at these locations with 1 second exposure time. The emission measure value at 1.1 R$_\odot$ during the CME was taken as 10$^{28.9}$ cm$^{-5}$ due to CME which was varied {up to 10$^{25}$ cm$^{-5}$} at the height of 1.5 R$_\odot$ with intermediate values interpolated using a third degree polynomial. It has been observed that the temperature of a CME can range from 10$^{5.5}$ to 10$^{6.5}$ {\citep{Susino2016ApJ}}, hence, an average temperature of 10$^{6.25}$ K was taken during simulation of this case. The coronagraph slits are kept in the reference position with the center of the Sun lying midway between slits 2 and 3. 

The synthetic spectra as will be observed by VELC corresponding to the 5303 \AA, 7892 \AA, and 10747 \AA\ {channels} are simulated and shown in Figure \ref{fig:cmecase}. Here, {the center of slit 4 (from the left)} is at a heliocentric distance of 1.11 R$_\odot$. In Figure \ref{fig:cmecase}(b,c,d) the expected synthetic spectra for the CME seen in the reference image is shown that will be observed by VELC at 5303 \AA, 7892 \AA, and 10747 \AA\ respectively. It can be seen that at the location of the CME, there is enhancement of peak intensity reflecting the increased electron density in the three channels of VELC. Comparing the peak electron counts in the presence of CME for the three channels with Table \ref{tab:50mic} it was found that there was an increment by $\sim$4, $\sim$5, and $\sim$150 times for the 5303 \AA, 7892 \AA, and 10747 \AA\ channels respectively. For the two visible channels the electron count is sufficiently below the full well capacity of the detectors even after adding the scattered photons. However, for the IR channel the counts are more than the full well detector capacity. This is due to larger pixel size of IR channel detector. This implies that even though exposure of more than 1 second could be set for visible channels, it should be less than a second for the IR channel for CME observations without saturating the detector. For this when the CME is detected onboard using the onboard CME detection logic \citep{Patel2018}, spectroscopic channel will be configured with a predetermined set of exposures. We plan to use the flag provided by the on-board CME detection algorithm for the continuum channel of VELC to change the IR observation to low gain mode with reduced exposure time for CME observations.  \\

\subsection{SNR Requirements for Specific Cases}
\subsubsection{Magnetometry}
{One of the aims of VELC is the measurement of coronal magnetic field using the IR channel of VELC. As the magnetic field in the active regions in the corona is of the order of few tens of Gauss \citep{Lin2000ApJ}, we synthesized the weakest Stokes profile, V, intensity for the IR channel keeping the EM and density same as { the} previous cases for slit-width of 50 $\mu$m and temperature of 10$^{6.2}$~K. The radial variation of the V signal  for magnetic field strength of 10 G is presented in Table \ref{tab:stokesv}. It could be noted that the V/I percentage of polarization for the instrument accounts to $\sim$0.13 \%. Taking in to account the Poisson noise at 1.1 R$_\odot$ the SNR will be $\approx$6. To { perform} the magnetic field measurement the SNR for Stokes-V should be in the order of $\sim$1000. { Acquiring this SNR requires integration times above one hour. Images with nominal exposure times will be recorded to avoid detector saturation and summed on the ground to produce sufficient SNR to provide critical measurements of the magnetic field in coronal active regions.}
\\}

\begin{table}[]
    \centering
    \begin{tabular}{ccccccc}
    \hline \\ 
        & Distance (R$_\odot$) & 1.1 & 1.2 & 1.3 & 1.4 & 1.5 \\ \hline
        & V & 35 & 13 & 7 & 4 & 3 \\
        & I & 26164 & 9894 & 5121 & 2937 & 1887 \\ 
        & \% of polarization (V/I) & 0.13194976 & 0.13195072 & 0.13195114 & 0.13195138 & 0.13195152 \\ \hline
    \end{tabular}
    \caption{Variation of Stokes-V signal radially in the VELC IR channel for 50 $\mu$m slit-width.}
    \label{tab:stokesv}
\end{table}

\subsection{Doppler Maps}
\begin{figure}
    \centering
    \centerline{\hspace*{0.05\textwidth}
               \includegraphics[width=0.5\textwidth,clip=]{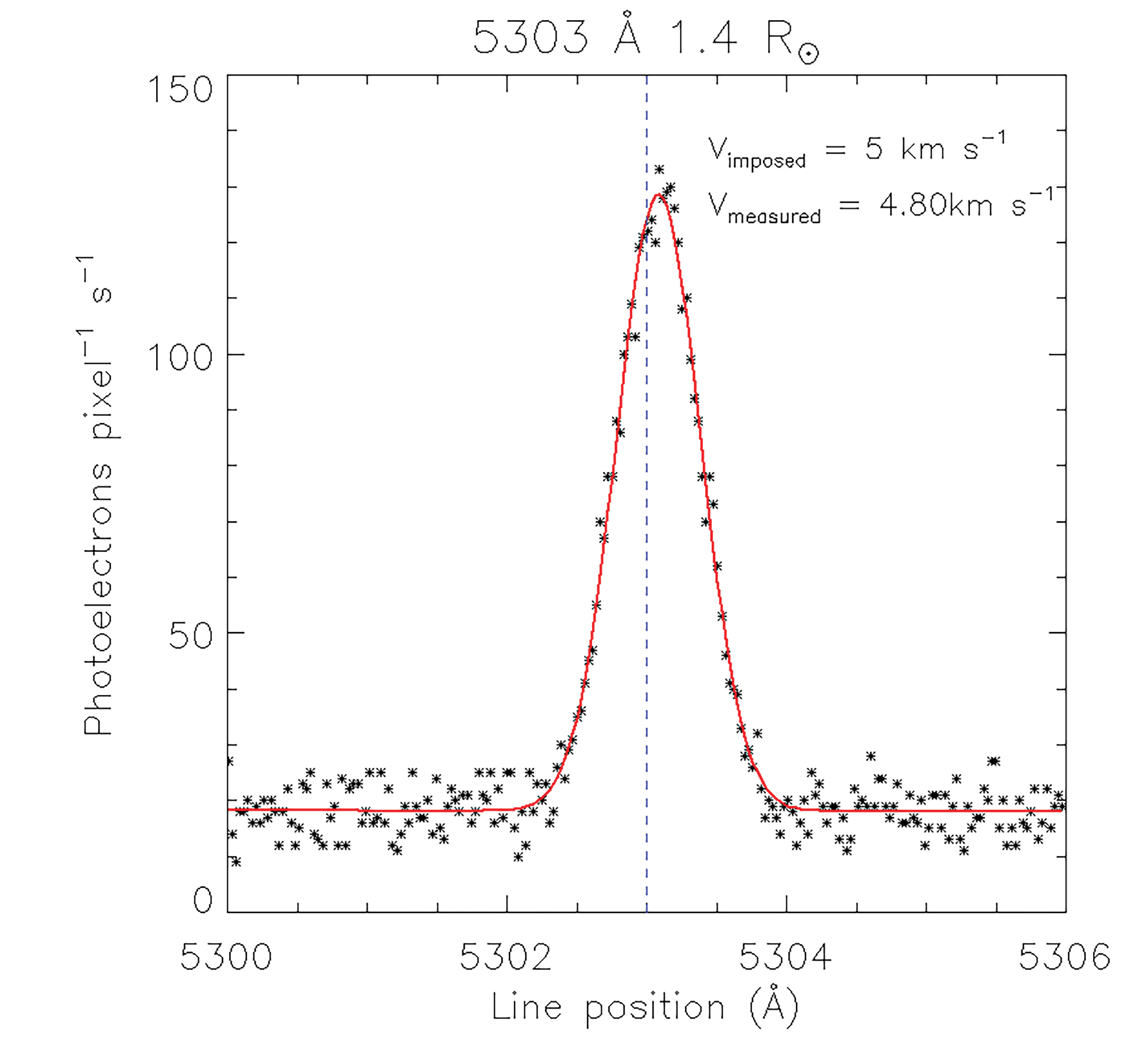}
               \hspace*{0.002\textwidth}
               \includegraphics[width=0.5\textwidth,clip=]{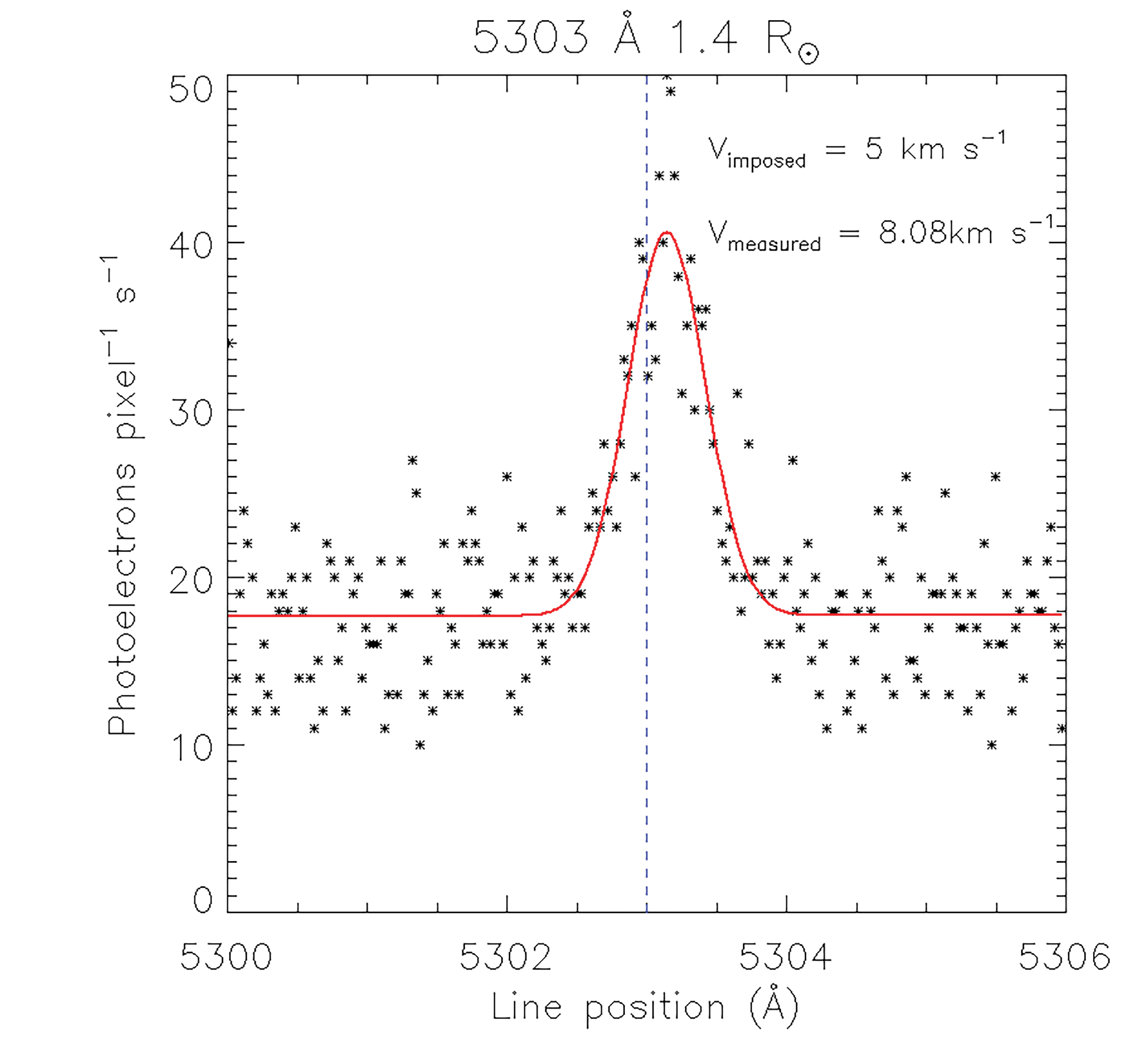}
              }
    \centerline{    
      \hspace{0.23\textwidth}  \color{black}{(a)}
      \hspace{0.5\textwidth}  \color{black}{(b)}
         \hfill}
      \vspace{0.01\textwidth}       
    \centerline{\hspace*{0.05\textwidth}
               \includegraphics[width=0.5\textwidth,clip=]{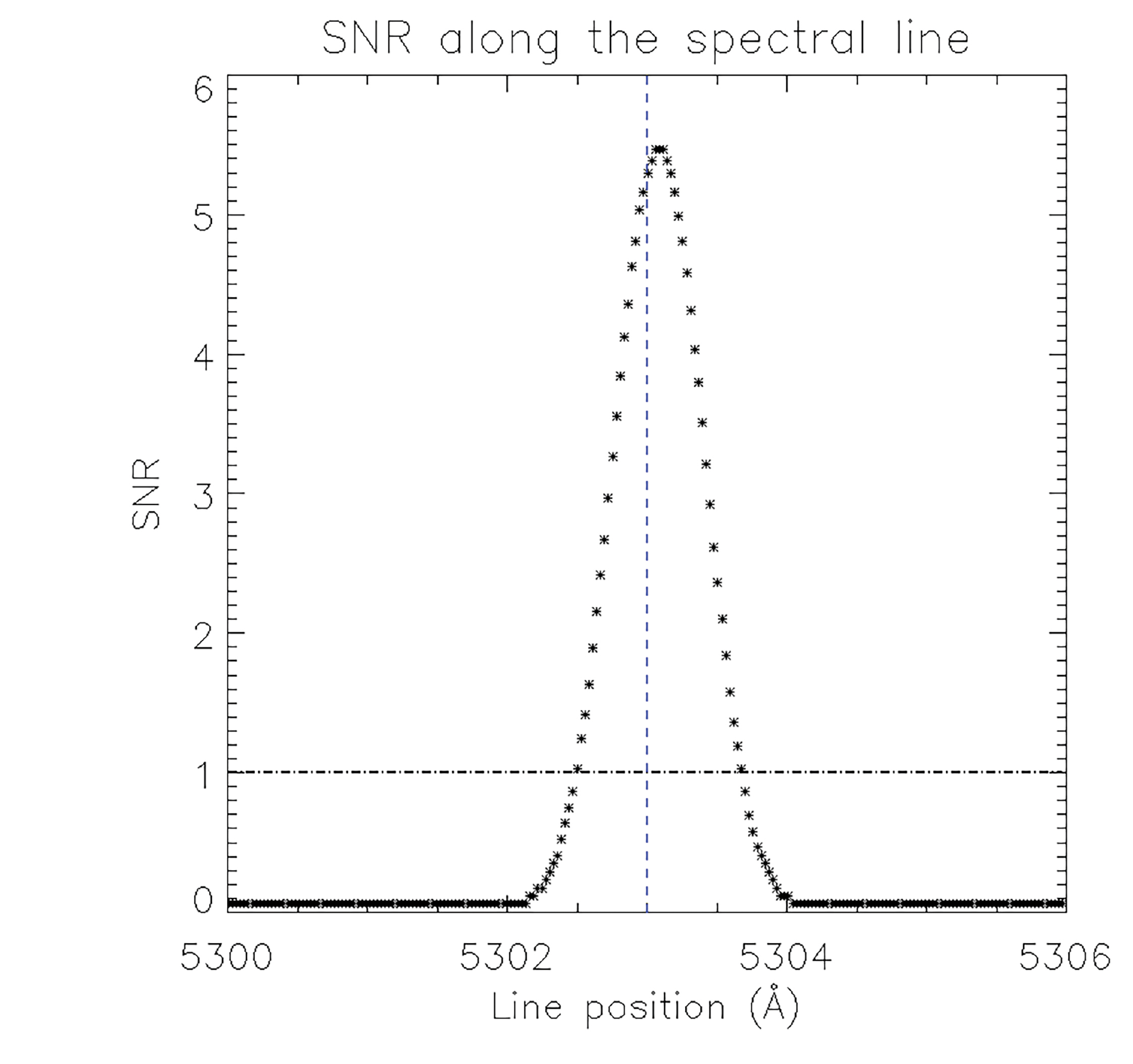}
               \hspace*{0.002\textwidth}
               \includegraphics[width=0.5\textwidth,clip=]{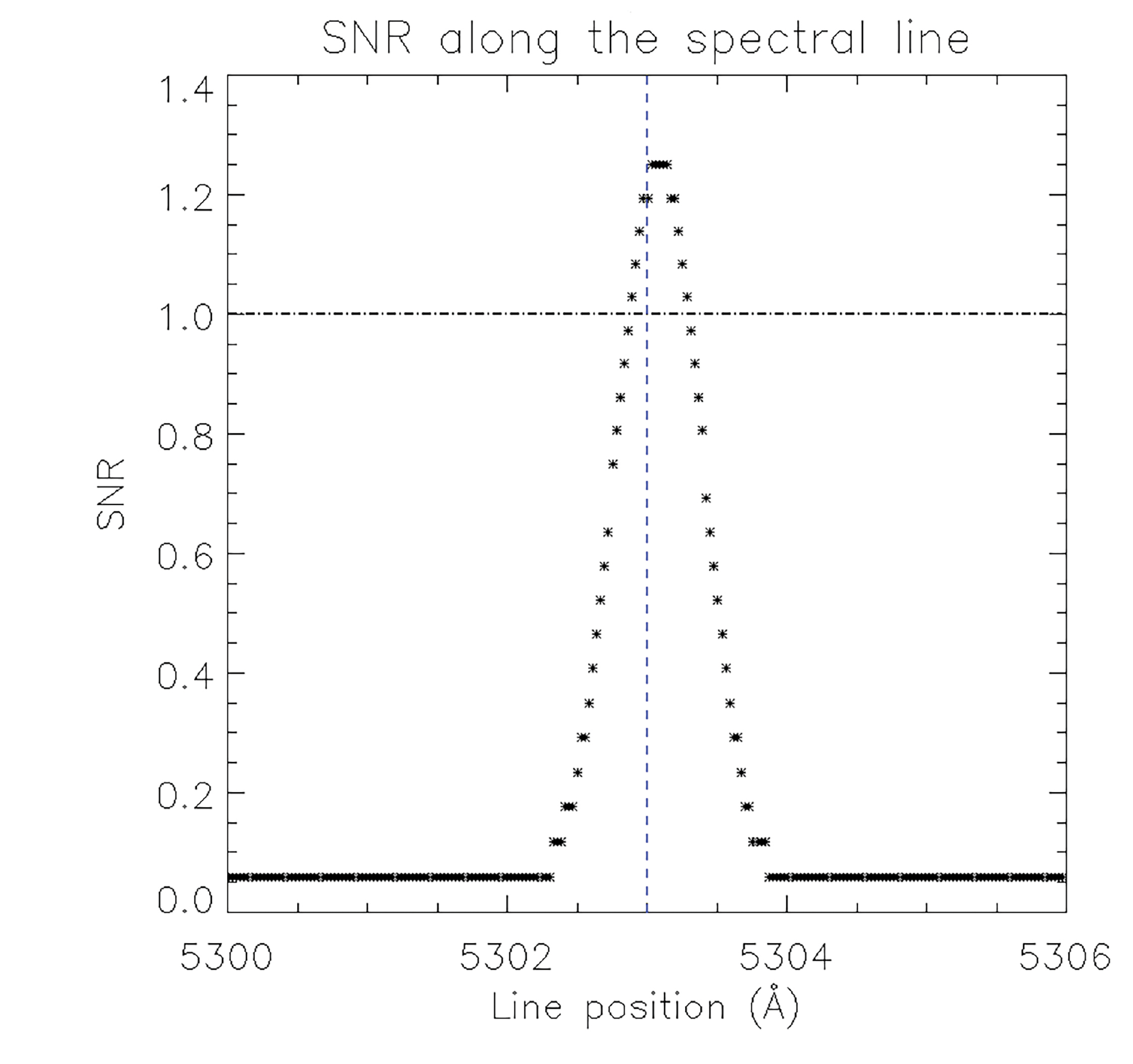}
              }
    \centerline{    
      \hspace{0.23\textwidth}  \color{black}{(c)}
      \hspace{0.5\textwidth}  \color{black}{(d)}
         \hfill}
    \caption{Synthetic green line with noise and doppler shift added (a) at 1.4 R$_\odot$, (b) at { 20\%} intensity of (a). The vertical dashed line shows the rest wavelength. (c), and (d) show SNR variation within the spectral line respectively for (a) and (b) where horizontal dotted-dashed line mark the SNR = 1.}
    \label{fig:doppler}
\end{figure}

{VELC will be used to produce doppler maps to study the { line-of-sight} plasma motions in the solar corona. This will also be helpful to study the turbulence generated in the corona. We used the synthetic spectra at { a} height of 1.4 R$_\odot$ where the SNR for 5303 \AA\ is $\approx$5 for the quiet-Sun case as mentioned in Table \ref{tab:50micpeak}. We then imposed a doppler velocity of 5 km s$^{-1}$ resulting in a shift of $\approx$88~m\AA\ for the line peak from the rest wavelength position. We added  the noise introduced in Section \ref{sec:results} randomly along the spectral line to mimic the near realistic observation. The final spectra was then fitted with a Gaussian profile and the doppler velocity was then measured. Figure \ref{fig:doppler}a and b shows the synthetic spectra at 1.4 R$_\odot$ and with { 20\%} counts of the prior. The vertical dashed line is the position of the rest wavelength for the spectra. The SNR variation within the line is shown in panels c and d of the same Figure. It could be seen that when the SNR at the peak of the spectra is $\approx$5, then a reliable doppler speed is measured in the synthetic data that is close to the imposed one. On the other hand when the SNR at the line peak is close to 1, then a significant deviation could be seen in the measured doppler speed (Figure \ref{fig:doppler}b). We found that SNR of { at least} 5 is good enough to measure the doppler speed as low as 5 km s$^{-1}$ using the green channel of VELC when all the noise sources are included in the line profile.}

\section{Summary and Discussions}
    \label{sec:summary}

VELC on-board Aditya-L1 will provide a unique opportunity to simultaneously image and perform spectroscopic observations of the inner solar corona in three visible and one IR pass-band.
It will be used for spectroscopic diagnostic of corona up to 1.5 R$_\odot$ using three emission lines, 5303 \AA, 7892 \AA, and 10747 \AA. It is necessary to simulate the performance of the instrument that will be useful for designing the observation plan after the launch. In this work we used synthetic spectral data to characterize the spectral channels of VELC for different solar conditions. We synthesized the spectra for the three channels using the CHIANTI atomic data base taking in to account the instrument characteristics. We also added the contribution of the instrument including the scattered intensity and detector noise to the synthesized spectra. The scattered intensity available for the continuum channel of the instrument was scaled for the spectral channel parameters. The final spectra was then analysed using the signal to noise calculated at the line center wavelength and at $\pm$0.5 \AA\ from the line center.

We simulated the synthetic spectra taking isothermal condition with average coronal temperature as 10$^{6.25}$ K for the three channels and estimated the SNR for the emission lines at coronal heights from 1.1 to 1.5~R$_\odot$. For the simulation the slit width was varied as 20 $\mu$m, 40 $\mu$m, and 60 $\mu$m. We found that on increasing the slit width the SNR at the peak intensity and at the wings increased for all the three channels.  It was identified that for the slit width of 60 $\mu$m, the detector for IR channels fills up to 77\% of its capacity with 500 ms exposure in high gain mode to be used for spectro-polarimetery mode. In order to keep a modest margin for this particular mode, the analysis was done taking slit width as 50 $\mu$m. It was found that 50 $\mu$m slit width leads to 65\% filling of the IR detector in spectro-polarimeter mode providing sufficient SNR at the same time. Thus, based on the requirement for this particular mode, we believe that the slit-width of 50 $\mu$m for the VELC spectral channels will be sufficient to study different  regions of the solar atmosphere.

It could be seen from Table \ref{tab:50mic} that the SNR for 7892 \AA\ is relatively poor as compared to the other two. This is because different lines have different formation temperatures. Therefore, we also studied the effect of different coronal plasma temperatures on the performance of the VELC spectral channels for the optimized slit-width. We synthesized the spectra for the three channels at 1.1 R$_\odot$ varying the temperature from log(T) = 6.0 to log(T) = 6.5 in steps of 0.1. We found that the SNR at line peak varies with change in temperature such that the maximum SNR  observed for 5303 \AA, 7892 \AA, and 10747 \AA\ was at log(T) of 6.3, 6.1, and 6.2 respectively. These temperatures are close to their line formation temperatures. We could observe a similar increase in SNR for the wings of the synthesized spectral lines. 
We analysed the instrument's performance with slit-width of 50 $\mu$m to synthesize the spectra with their line formation temperatures. The SNR was then estimated at the line peak and the wings for coronal heights from 1.1 to 1.5 R$_\odot$. From the results of this analysis presented in Table \ref{tab:50micpeak}, it could be observed that there is sufficient SNR in all the channels for the line peak as well as the wings which decreases at larger heights. The IR channel has very good SNR even at larger heights due to its bigger pixel size as compared to the other two.

As these analysis were based on isothermal coronal conditions, we also used a MCMC simulation taking a Gaussian distribution of temperature peaking at 10$^{6.1}$ K having width of 10$^{0.3}$ K to estimate the expected signal at different heights in the VELC FOV. We found that there is spread in the estimated counts based on the temperature distribution with a cluster obtained corresponding to the temperatures close to the line formation temperature for individual channels (Figure \ref{fig:dist_height}).
Overall analysis shows that for the optimized slit-width of 50 $\mu$m and considering the peak line formation temperatures of each line, reliable signal could be obtained for all the channels which could be used for analysis. For the larger heights when the SNR becomes $\leq$5, then pixel binning could be considered to enhance the signal with respect to the background. There is also an option to increase the exposure time without saturating the detector which could also boost the signal. For the study of very fast transients requiring short exposure times, the observations could be taken at high cadence. Such short exposure frames could be further binned to increase the SNR. 

We also performed a study of a CME case approximating the enhanced electron density due to a CME of 2016-01-01 from KCor images which has similar FOV as VELC. We found that the visible spectral channels of VELC could be operated for CME observations with exposure time more than 1 s whereas for IR channel will be operated in low gain mode following the trigger provided by on-board CME detection algorithm with reduced exposure time. More CME cases could be analysed in future with different enhancements that could help in planning for spectral diagnostics of CMEs using VELC. 
{We also considered two specific science cases', magnetometry and doppler mapping, SNR requirement from VELC point of view. For the measurement of magnetic field using the IR channel we calculated the Stokes-V intensity thereby estimating the minimum integration time required. For the doppler mapping, we used the green line at height of 1.4 R$_\odot$ where SNR $\approx$5. We also considered a case where the intensity of the line is 20 \% of the above mentioned case. For the two cases we added the noise and a doppler speed which was measure later by fitting a Gaussian. We found that a minimum SNR of 5 will be required for doppler mapping using VELC.}

{It should be noted that increasing the slit-width results in broadening of the spectral lines and hence decreasing the spectral resolution. \citet{Singh2019} studied the effect of increasing the slit-width for the three spectral channels of VELC. Their analysis also indicates that the slit-width should be increased to enhance the SNR at the same time optimising for the spectral resolution to meet the science requirements. These estimates will be helpful to identify the expected observed spectra including the instrument contribution which could be de-convolved during the processing to get the true line profile. The effect of spacecraft drift and jitter as studied by \citet{ranganathan2019polarimeter} has not be considered for the studies presented here but will be included in our future work. }
Such complete spectral information will be helpful for the data pipeline development and extracting the signal from the background

In this work we have presented a few of the solar conditions. A similar study could be extended to other solar features such as loops, plumes, coronal holes etc. which results in different ambient coronal conditions. Such extensive studies covering different cases could help in preparing the optimised plan for maximising the science output from the instrument. It should be noted that this study could also be extended for future missions which include spectrographs for preparing the target science cases that could be addressed with the instrument capabilities.

\section*{Conflict of Interest Statement}

The authors declare that the research was conducted in the absence of any commercial or financial relationships that could be construed as a potential conflict of interest.

\section*{Author Contributions}

MA, VM, and AKS generated the synthetic spectra for different solar conditions using CHIANTI database taking instrument parameters. RP and VM converted the synthesized spectra to the instrument's synthetic observations. RP carried on the estimations and prepared the manuscript. KS and DB planned the analysis from the observation point of view. VP provided the essential inputs to analyse the results. All authors took part in the discussion.


\section*{Acknowledgments}
We would like to acknowledge IIA, ISRO and ARIES to provide necessary facilities and computation requirements. We thank the VELC instrument team to provide the instrument parameters when required. We also thank the CHIANTI team for making the database available. RP and AKS would like to thank Samrat Sen for discussions regarding the analysis. 
RP, MA and VM are supported by DST. CHIANTI is a collaborative project involving the University of Cambridge (UK), the NASA Goddard Space Flight Center (USA), the George Mason University (GMU, USA) and the University of Michigan (USA).



\bibliographystyle{frontiersinSCNS_ENG_HUMS} 

\end{document}